\documentclass[twoside]{Sciexplor}

\colorlet{orange}{black}
\articleset{
    firstrhead = \\DOI: XXXX\vskip-1ex,
    articletype= Perspective,
    title = {Reframing Human–Robot Interaction Through Extended Reality: Unlocking Safer, Smarter, and More Empathic Interactions with Virtual Robots and Foundation Models},
    author = \textcolor{orange}{Yuchong Zhang, Yong Ma, Danica Kragic},
    affliction = 
      {
        \textsuperscript{1} Division of Robotics, Perception, and Learning, KTH Royal Institute of Technology, Stockholm, Sweden.\\
        \textsuperscript{2} Department of Clinical Medicine, University of Bergen, Bergen, Norway.
      },
    correspond = {Dr. Yuchong Zhang, Division of Robotics, Perception, and Learning, KTH Royal Institute of Technology, Stockholm, Sweden. Email: \url{yuchongz@kth.se}},
    received = date month year, 
    accepted = unknown,
    published = unknown,
    rhead = Vol 1 Issue 1,
    lhead = Journal Name,
}

\begin{document}

\maketitle

\linespread{1.2}

\begin{abstract}
This perspective reframes human–robot interaction (HRI) through extended reality (XR), arguing that virtual robots powered by large foundation models (FMs) can serve as cognitively grounded, empathic agents. Unlike physical robots, XR-native agents are unbound by hardware constraints and can be instantiated, adapted, and scaled on demand, while still affording embodiment and co-presence. We synthesize work across XR, HRI, and cognitive AI to show how such agents can support safety-critical scenarios, socially and cognitively empathic interaction across domains, and outreaching physical capabilities with XR and AI integration. We then discuss how multimodal large FMs (e.g., large language model, large vision model, and vision-language model) enable context-aware reasoning, affect-sensitive situations, and long-term adaptation, positioning virtual robots as cognitive and empathic mediators rather than mere simulation assets. At the same time, we highlight challenges and potential risks, including overtrust, cultural and representational bias, privacy concerns around biometric sensing, and data governnance and transparency. The paper concludes by outlining a research agenda for human-centered, ethically grounded XR agents -- emphasizing multi-layered evaluation frameworks, multi-user ecosystems, mixed virtual–physical embodiment, and societal and ethical design practices to envision XR-based virtual agents powered by FMs as reshaping future HRI into a more efficient and adaptive paradigm. \\

\textbf{Keywords:} Human-robot interaction, extended reality, large foundation models, cognitively empathic interaction
\end{abstract}

\section{Introduction}
As the prosperity of robotics during past years, intelligent robots have been more integrated into everyday lives \cite{xia2024shaping}. In particular, human–robot interaction (HRI) became a crucial field by intertwining humans with robots in human-centric environments \cite{zhang2024vision}. However, HRI has traditionally been rooted in the constraints of the physical world, prioritizing mechanical safety, real-time control, and spatial coordination between humans and embodied machines. These systems have yielded significant advances in domains such as manufacturing, logistics, and assistive robotics \cite{ringwald2023should,karabegovic2015application,schmitt2018soft,zhang2025mind}, yet they remain bounded by several practical limitations, such as the cost of hardware, the risks of human–machine co-location, and the inflexibility of physical deployment in rapidly dynamic or hazardous environments \cite{trevelyan2016robotics,willemse2017affective}. \\

\begin{figure}[!t]
    \centering
    \includegraphics[width=\textwidth]{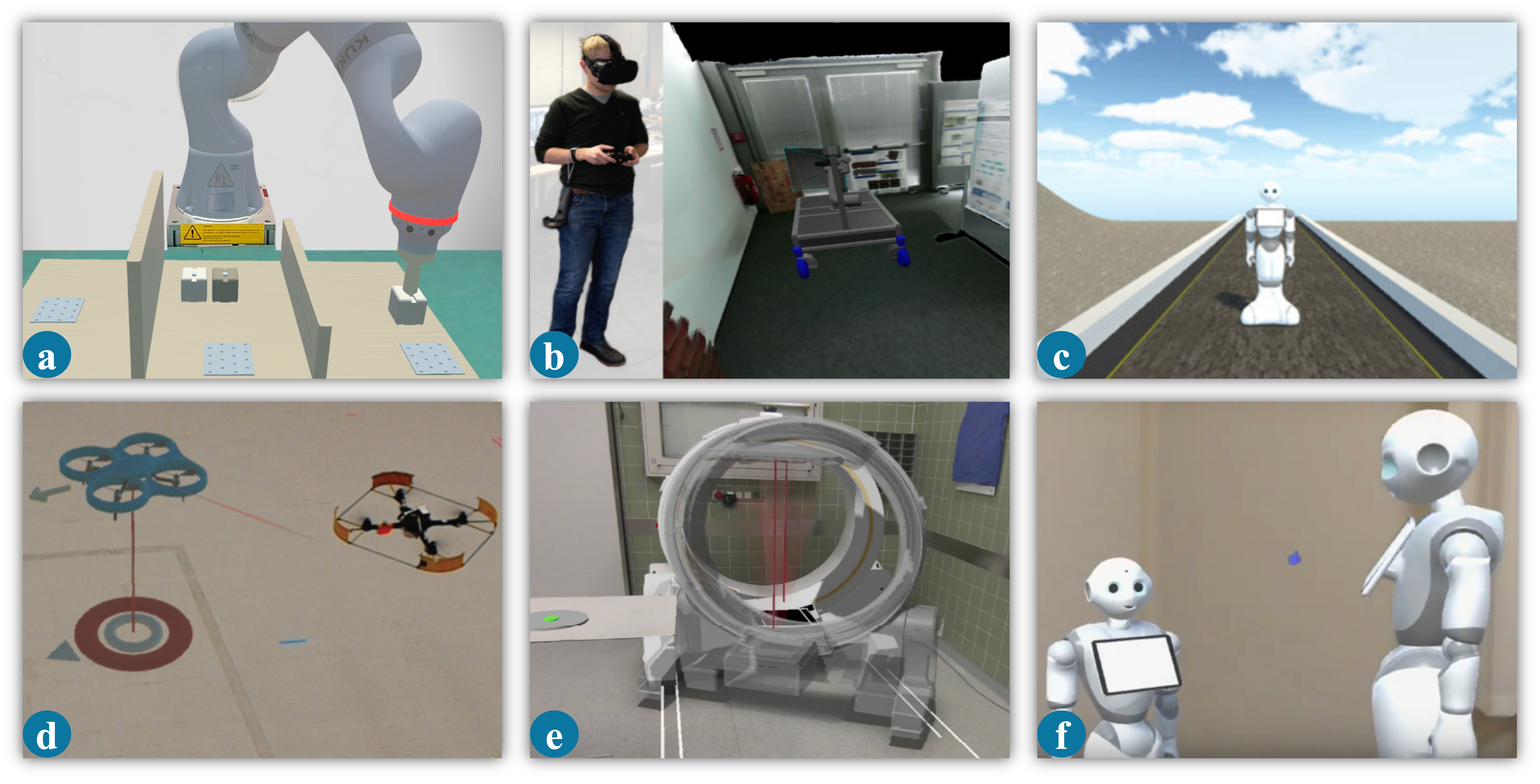}
    \caption{Several examples of human-virtual robots interaction powered by XR. $a$: virtual robot arm in VR \cite{mielke2025virtual}; $b:$ virtual mobile robot in VR \cite{stotko2019vr}; $c:$ virtual humanoid robot in VR \cite{li2019comparing}; $d:$ virtual flying robot (drone) in AR/MR \cite{walker2019robot}; $e:$ virtual medical robot in AR/MR \cite{plumer2024xr}; $f:$ virtual humanoid robots in AR/MR \cite{peters2018towards_virtual}.}
    \label{example}
\end{figure}

Meanwhile, extended reality (XR) -- an umbrella term encompassing virtual reality (VR), augmented reality (AR), and mixed reality (MR) -- has matured into a powerful paradigm for immersive interactions, multimodal interaction, and elevated cognition in the field of general human-computer interaction (HCI) over the past decades \cite{zhang2023see}. By incorporating virtual information visualized either in complete virtual world or real-world, XR offers not only visual-spatial fidelity but also the capacity for dynamic feedback, co-presence, and more engaged interaction in many domains \cite{kim2022effects,lee2025conceptual}. There has been massive research that has delved into employing XR techniques to facilitate advanced HRI by bringing virtual setting, which makes it uniquely well-situated to complement and extend the principles of HRI beyond the physical workspace (Figure ~\ref{example}).\\

In parallel, we are witnessing an unprecedented shift in artificial intelligence (AI) through the development of large foundation models (FMs) \cite{awais2025foundation} -- particularly large language models (LLMs), large vision models (LVMs), and vision–language models (VLMs) -- that endow agents with context-aware reasoning, conversational competence, and multimodal perception \cite{lu2023mathvista,li2024multimodal,ma2024understanding}. When integrated with HRI, these models enable the creation of more smooth interactions between humans and robots due to the algorithmic advancements, for example, in communicating robot intent or learning from human feedback \cite{lin2020review,kupcsik2017learning}. When coupled with XR, the involvement of virtual elements would lead to another potential approach for interaction \cite{nowak2021augmented}. For instance, intelligent virtual robots -- agents that inhabit immersive spaces -- can understand user intent and adaptively respond through natural language, gaze, gesture, or other various inputs \cite{park2021hands,qi2024computer,lynch2023interactive,ruhland2015review,zhang2024human}. \\

The virtual robots are more than plain visualizations or surrogates \cite{plumer2024xr,walker2019robot} of corresponding physical systems. Besides serving as the virtual counterparts, they also represent a new category of cognitive agents -- capable of participating in collaborative tasks, simulating empathic engagement, and supporting experiential learning without the material constraints of conventional robotics. In industrial training scenarios, for example, XR-based robots have been shown to improve domain task performance, user experience, and procedural safety awareness \cite{dianatfar2024virtual,karpichev2024extended}. In specific therapeutic and educational contexts, socially responsive virtual agents powered by LLMs are already being explored as scalable tools for mental health interventions, language acquisition, and neurodiverse engagement \cite{zhang2023large,wang2024large}. This evolution of immersive media technology and cognitive AI, alongside the intersection between them, encourage us to reflect: while the traditional robotic paradigm emphasized physical manipulation, like grasping, moving, and executing, more emerging paradigms emphasize co-presence and the co-construction of tasks through immersive interactions \cite{duguleana2011evaluating,zhang2023playing} with the engagement of virtual elements \cite{weistroffer2014assessing,bayro2022subjective,zhang2022initial}. This shift reframes HRI from a primarily mechanical interface to a socio-cognitive encounter mediated by technology. \\

By witnessing the current landscape of interacting with virtual robots through the lens of XR -- native virtual robots -- autonomous, adaptive agents that reside entirely within immersive environments, we argue that in the future, such virtual agentic loops offer not only practical benefits—safety, scalability, and cognitive adaptability—but also open new directions for understanding human–machine interaction as a fundamentally relational process. We explore how humans engage with these virtual robots as social, instructional, and empathic partners across diverse domains, such industrial safety training, education, therapeutic interventions, collaborative design, etc. By narrowing the scope to XR-native agents, we are able to examine the unique affordances and challenges inherent in this paradigm: embodiment without physical form, presence without mass, and cognition unconstrained by mechanical limitations. This framing also enables us to explore how emerging AI models (e.g., LLM, VLM) -- can endow these agents with contextual awareness, responsiveness, and the capacity for long-term adaptation. By synthesizing developments across XR, HRI, and large-scale FMs, we articulate a research agenda for the next generation of  empathic and ethically grounded virtual robotics. \\

\subsection*{Positioning}
\begin{figure}[!t]
    \centering
    \includegraphics[width=\textwidth]{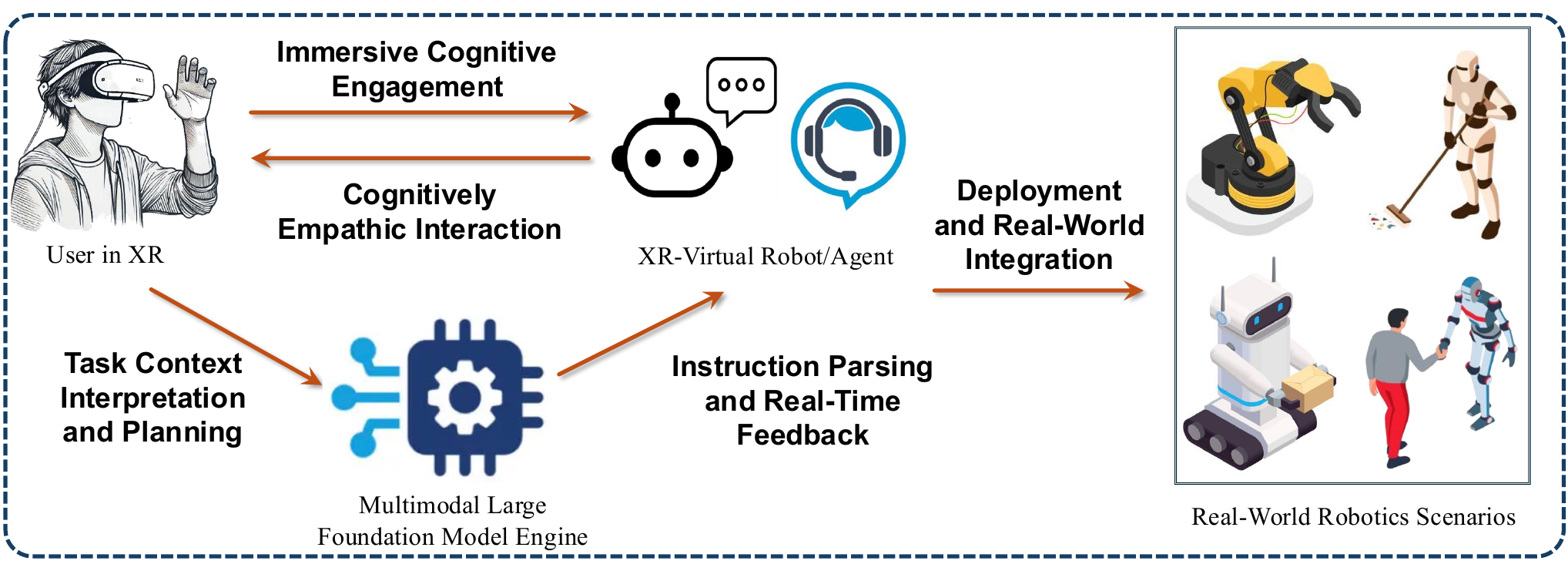}
    \caption{XR-enhanced cognitive and empathic interaction flow between human-virtual robots supported by large FMs.}
    \label{concep}
\end{figure}

This paper investigates the interaction between humans and intelligent virtual robots that exist entirely within XR environments. We regard virtual robots as embodied, agentic entities -- either humanoid or non-humanoid in form -- rendered in immersive digital environments with perception, reasoning, and action capabilities via AI. These agents are native to XR: they operate independently of physical actuators or hardware proxies and are designed for real-time, situated interaction within virtual contexts. Our focus outreaches from traditional robotic paradigms that employ XR technique as a visualization tool or user interface overlay. Apart from physical robots augmented by XR for specific task-oriented teleoperation or collaborative support, we concentrate on fully virtual robots capable of functioning as standalone agents with roles, such as instructors, collaborators, companions, or cognitive interfaces in immersive, spatial environments actuated by intelligent large FMs. A conceptual diagram of XR-empowered human-virtual robot with FMs is illustrated in Figure ~\ref{concep}. \\

\section{Current Research Landscape}
The intersection of XR and HRI has witnessed rapid growth in recent years, driven by advances in immersive technologies, intelligent agent architectures, and the democratization of AI models. Across both industrial and academic settings, XR platforms are increasingly being used to simulate robot behavior, mediate collaboration, and enable safer, more personalized forms of interaction.

\subsection{Safety-Aware XR for Human–Robot Collaboration}
The use of XR  -- mostly VR and AR -- has gained considerable traction in enabling safety-aware human–robot collaboration (HRC), especially in industrial, high-risk, or mission-critical domains. XR systems enable the development of virtual elements or digital twins of physical robots that support users in visualizing robot behaviors, simulating hazardous tasks, and reducing physical risks during planning and training phases.

\subsubsection*{Visualizing Robot Behavior for Hazardousness}
Multiple studies emphasize the power of XR to make robot behavior more transparent and predictable through visual overlays and real-time cues. Enayati et al.~\cite{karpichev2024extended} propose a human-in-the-loop XR system to visualize robot trajectories and operational zones. Similarly, MR interfaces evaluated by~\cite{san2025mixed} use audio–visual cues to render dynamic hazard zones, increasing the user awareness. Works such as~\cite{bolano2019transparent_virtual,walker2018communicating,walker2019robot} explore AR projections or visualizations of robot motion intent or predicted paths, improving perceived safety, interaction naturalness, and reducing ambiguity during close-range collaborations with robots. Digital twins are increasingly used to visualize risk and monitor remote systems. As shown in works like~\cite{malik2021digital,xie2024new,choi2022xr}, synchronizing virtual and physical robots allows users to anticipate unsafe interactions and maintain awareness during shared-space operations. There are several literature reviews like~\cite{wang2025systematic,pan2024integrating,green2007human} which further highlight the role of digital twins and virtual agents in visualizing context and enhancing transparency.

\subsubsection*{Training and Planning with Simulated XR}
XR-based training tools have demonstrated strong potential in improving procedural safety and knowledge retention in HRI~\cite{Duguleana2011eva,Higgins2021_virtual,Murnane2021_virtual}. Dianatfar et al.~\cite{dianatfar2024virtual} show how VR welding simulations lead to increased user confidence and task proficiency. Similar systems~\cite{arntz2024enhancing} offer immersive environments where users were engaged with simplified navigation alongside robot behavior. Surve et al. \cite{surve2024unrealthasc} enabled HRI training to be conducted in photorealistic simulated environments, using XR wearables with virtual underwater robots, before deployment in real-world scenarios where safety could be at risk. Virtual laboratories, allowing for reproducible and controlled studies of HRC under complex psychological and ergonomic conditions~\cite{bustamante2021armsym,villani2018use}, were developed for enhanced interaction with safety. For example,~\cite{fratczak2019understanding} examines how robot motion affects human posture and stress using VR-based observation, while~\cite{yamamoto2012augmented} combined AR with haptic feedback to reduce cognitive load during robot-assisted surgical situations. For task planning, some systems help reduce planning errors and allow iterative testing without physical execution, such as \cite{fang2014novel} which provided AR interfaces to preview and adjust robotic motion paths or \cite{hernandez2020increasing_virtual} which offered virtual robot in AR for high-level requesting before accurate motion planning. For industrial assembly contexts, AR-based tools assist in human-robot alignment and workflow optimization~\cite{kousi2019enabling,michalos2016augmented} for improved task efficacy.

\subsubsection*{Remote and Distributed Safety-aware HRI}
Remote interaction with robots or robotic teleoperation through XR is also gaining prominence~\cite{Milgram1993,Maly2016,Ostanin02122021,fang2022brain}. Studies such as~\cite{guhl2018enabling,wang2024robotic} combine XR and networked control for distributed robot supervision. Through immersive visualizations and overlays from VR and AR, users retain situational awareness and control without excessive direct physical presence in adaptive teleoperation, where intuitive visual, audio, or mid-air feedback~\cite{bischoff2004perspectives,hedayati2018improving} are accompanied.

\subsubsection*{Adaptive, Accessible, and Inclusive XR-HRI}
Despite broad applicability, many XR-HRI systems still lack adaptability or emotional responsiveness. Some reviews such as~\cite{wang2025systematic,pan2024integrating} highlight that most implementations are tailored to single-user setups or rigid scenarios. Recent works have begun to address these gaps through individualized or personalized  safety concerns and adaptive communication modeling~\cite{choi2022integrated,Li2023assist}. XR is also proving to be a low-cost, accessible means of robot interaction. Mobile smartphone-based AR interfaces~\cite{chacko2019augmented} enable seamless robot manipulation in unconstrained settings. Wearable AR provides hands-free guidance, visual feedback, and improved interaction with higher effectiveness in dynamic environments~\cite{ostanin2019interactive,manring2019augmented}. \\

\textbf{In summary}, XR-based robot systems using additional visualized virtual elements contribute significantly to safety-aware HRI/HRC by enabling informative visualizations, safe training environments, and more intuitive interactions. Moving forward, research should continue to explore adaptive, intelligent, and inclusive XR interfaces to fully realize the potential of virtual robotic collaboration in complex and dynamic settings.

\subsection{Cognitive and Empathic Virtual Robots: Large FMs with XR}
The rise of large FMs during past years is enabling a new generation of cognitive and empathic norms of both advanced robotic and HRI loops. No longer limited to passive interfaces or scripted responses, virtual agents embedded in XR environments can now engage in contextual reasoning, adapt behavior in real time, and respond empathetically to users due to generative AI.

\subsubsection*{FMs as Cognitive Engines in XR-HRI}
Numerous recently developed platforms and frameworks integrate LLMs into XR to support spatially and semantically grounded communications with agents. CUIfy~\cite{buldu2025cuify} demonstrates an open-source pipeline for embedding LLM-powered conversational agents in Unity XR environments, minimizing latencies during interactions. Konenkov et al.~\cite{konenkov2024vr} combine VR with VLMs, enabling reduction in task completion time but elevation in user comfort. Afzal et al.~\cite{afzal2025next} highlighted that LLMs with XR would facilitate situational awareness while underlining ethical considerations. Li et al.~\cite{li2024vision} showed that foundational VLMs can make robotic control more flexible while being a cost-effective and easy-to-use solution. Synthetically,~\cite{Lakhnati2024exploring} explored a GPT-based human-robot teaching through VR settings, which proved the functionality of simulated virtual robots powered by FMs.

\subsubsection*{Empathic and Emotionally Aware Virtual Robots}
Empathy-focused HRI is a growing priority in XR systems. The review by~\cite{ZHANG2023100131} detailed how LLMs facilitate emotion recognition, conversational nuance, and contextual understanding in HRI. Social cues, such as haptics are also simulated in XR. For example,~\cite{wang2011handshake} presented a haptic-enhanced VR handshake system with visual cues to enhance social presence for robotic partnership. ~\cite{bustamante2021armsym} enabled empathetic testing of prosthetic interaction through a safe virtual laboratory. Understanding user stress, posture, and motion in XR has also been studied for adaptive and empathic robot control and interaction~\cite{fratczak2019understanding,villani2018use}. Notably, some AR-related works in HRI are now also augmented with empathy-aware overlays. For example, Wang et al. explored user preference of embodied virtual agents in AR with different design propositions~\cite{wang2019exploring}. Systems like~\cite{bolano2019transparent_virtual,walker2018communicating} enhanced user comfort, trust, and efficiency by previewing robot motion intent through AR. Empathic engagement in high-risk environments can be further enabled by XR agents that visualize shared intentions~\cite{green2007human,suzuki2022augmented}.

\subsubsection*{Human-centered and Multimodal Interaction}
To support human-centered core, while retaining trustworthy and intuitive virtual agents in XR-based HRI, several works proposed distinct taxonomies and design principles. These include the virtual element design taxonomy~\cite{walker2023virtual}, the dedicated AR-HRI classification~\cite{suzuki2022augmented}, and the key characteristics of robots~\cite{groechel2022tool}, which help researchers build user-centered, expressive, and empathic agentic interactions. Design-oriented XR systems like~\cite{peters2018towards_virtual,wang2015intelligent,szafir2019mediating,qiu2020human} explore proxemics, social communication, and information exchange, facilitating emotionally intelligent design of HRI systems. In terms of interaction modalities, gesture and non-verbal input are critical components of empathic HRI. Gesture recognition in VR~\cite{sabbella2023virtual} and multimodal input~\cite{Arntz2020_virtual} via wearable MR~\cite{park2021hands} enable virtual robots to interpret user intent and adjust behavior dynamically. \\

\textbf{In summary}, human-centered, cognitive, and empathic virtual robots especially powered by FMs are transforming XR into a shared-platform for more adaptive and emotionally aware HRI. For instance, through LLM integration, multimodal feedback, and empathic design, virtual agents now serve not only as tools, but as collaborators capable of sensing and understanding complex human needs.

\begin{figure}[!t]
    \centering
    \includegraphics[width=\textwidth]{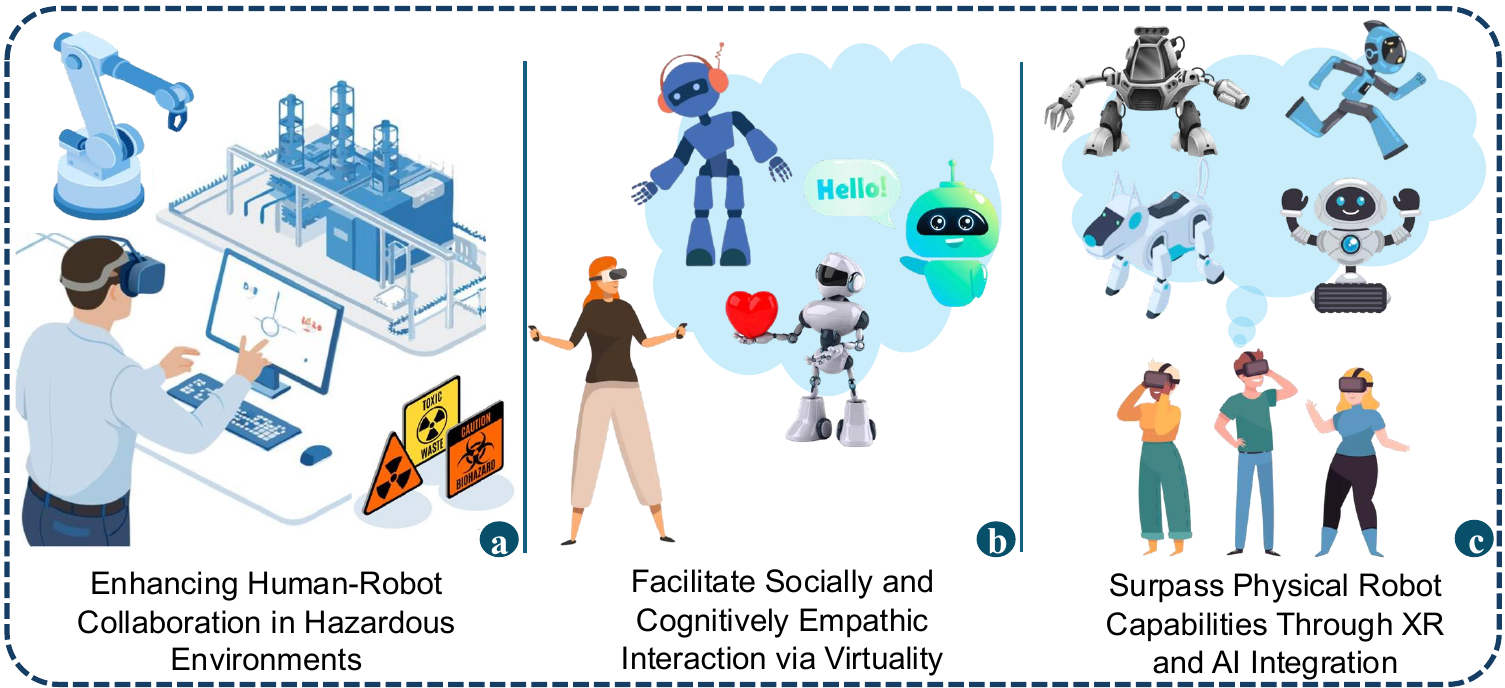}
    \caption{Future scenes with XR-powered virtual robots.}
    \label{future}
\end{figure}

\section{Future Scenarios of XR-Enhanced Virtual Robot Interaction}
By revisiting the current research and application status of XR-based HRI as well as the surge of large foundational models, the next generation of XR-enabled virtual robots will go beyond merely simulation and visualization \cite{zhang2021supporting}. These virtual robots are envisioned to be developed with agentic empathy \cite{paiva2017empathy} to become different roles, such as adaptive collaborators and companions capable of reasoning, teaching, and responding to user needs across a range of real-world tasks. We outline here three forward-looking domains where virtual robots could deliver transformative impact, while the three corresponding scenarios are shown in Figure ~\ref{future}.

\subsection{Enhancing Industrial Collaboration in Hazardous Environments}
Industrial domains such as manufacturing, construction, biochemical processing, and emergency cases frequently involve hazardous conditions that challenge conventional human–robot co-presence \cite{zhao2003toward,weistroffer2014assessing}. Safety concerns, limited visibility, high-risk machinery, and unstable environments often preclude direct interaction with physical robots. XR technologies offer a transformative alternative: by simulating immersive digital twins of real-world operations, workers can engage with virtual counterparts of industrial robots to rehearse procedures, calibrate workflows, and develop collaborative pipelines with mitigating physical risks. \\

Within these environments, users can interactively visualize safety-critical zones, explore dynamic task trajectories, and manipulate virtual robots with high fidelity to their real-world equivalents \cite{szczurek2023multimodal}. VR is particularly effective for simulating chaotic or unpredictable scenarios where physical testing is impractical. In such settings, virtual robots -- programmed with behaviors modeled after their physical counterparts -- can provide users with realistic and repeatable task rehearsals that improve coordination and situational awareness \cite{simaan2015intelligent,roldan2017multi}. \\

AR or MR expand these capabilities into real-time field operations \cite{zhang2023industrial}. For example, through optical see-through head-mounted displays (HMDs), users can view superimposed virtual robots aligned with their physical environments, enabling spatially accurate remote teleoperation \cite{van2024puppeteer,zhang2025llm,zhang2025multimodal}. This is especially valuable in mobile robotics designated into the on-site hazardousness-aware situations, where AR overlays allow for user interaction and monitoring at a safe distance, while maintaining functional continuity with on-site deployments. \\

By integrating real-time sensory feedback and AI-driven reasoning (e.g., via LLMs), these virtual agents adaptively interpret human intent and respond to dynamic conditions. More than functional simulators, they serve as communicative intermediaries -- visualizing robot intent, forecasting motion paths, and modeling shared control frameworks. Such transparency builds user trust, reduces cognitive load, and fosters safer, more effective collaboration during both training and live operations \cite{kim2024understanding,kawaharazuka2024real,costa2022augmented}. As such, XR-enabled virtual robots act as critical mediators in hazardous environments, bridging the gap between simulation and safer deployment. \\

\subsection{Empowering Socially and Cognitively Empathic Interaction via Virtual Humanoid Robots}
In social, healthcare, and educational contexts, humanoid robots are increasingly deployed to support therapy, rehabilitation, companionship, and interactive learning \cite{robins2005robotic,shamsuddin2012initial,mohebbi2020human,leite2013influence}. However in numerous cases, physical platforms remain expensive, difficult to adapt, and are often limited in their range of emotional expression. Virtual humanoid robots in XR alleviate many of these constraints by existing as fully programmable, reconfigurable avatars that can be tailored to individual users and scenarios. Within immersive environments, these agents can inhabit shared spaces with users, maintain functionality, modulate input signals such as voice and posture, and enact subtle social cues, practically achieving a richer empathic interaction than many current physical embodiments \cite{baytas2019design}. \\

The integration of large FMs further amplifies these capabilities. By combining LLMs, LVMs, and VLMs, virtual agents can interpret multimodal input -- speech, gaze, audios, facial expression, and gesture -- and respond in a manner that is contextually appropriate and emotionally attuned. In VR, fully immersive environments populated with such agents can be used to rehearse or remaster challenging social situations, practice turn-taking with more empathy \cite{shin2018empathy,rueda2020virtual}. The game-like learning experiences in VR with virtual robots can reward mutual collaboration and facilitate curiosity, improving the overall learning efficiency in educational cases \cite{lv2022application}. In AR, virtual companionship robots can be overlaid onto ordinary environments, functioning as “always-available” conversational partners who accompany users in need through mentally therapeutic exercises \cite{zhang2025personalizing}. By enabling empathic conversations with target users, without the physical constraints of moving and operating hardware, virtual empathic agents can more effectively facilitate communicative interaction \cite{nakazawa2023augmented,billinghurst2014using,ma2023emotion}. \\

These virtual companions are particularly well-suited for supporting heterogeneous user groups, ranging from language learners who benefit from personalized tutoring to patients undergoing exposure therapy for anxiety or mental trauma. Because their forms and behavior are reconfigurable, virtual robots can be adapted to different cultures, age groups, and appearance preferences without redesigning hardware. Their consistent and programmable empathy make them promising supports for vulnerable populations, including children on the autism phase or older adults experiencing cognitive decline. As AI-driven virtual agents continue to advance in empathy modeling and affective reasoning \cite{sorin2024large,hasan2023sapien,shih2024empathy}, XR systems may increasingly sustain relationships that are practically meaningful, while dynamically adapting to users’ emotional states and evolving alongside therapeutic or educational goals.

\begin{figure}[!t]
    \centering
    \includegraphics[width=\textwidth]{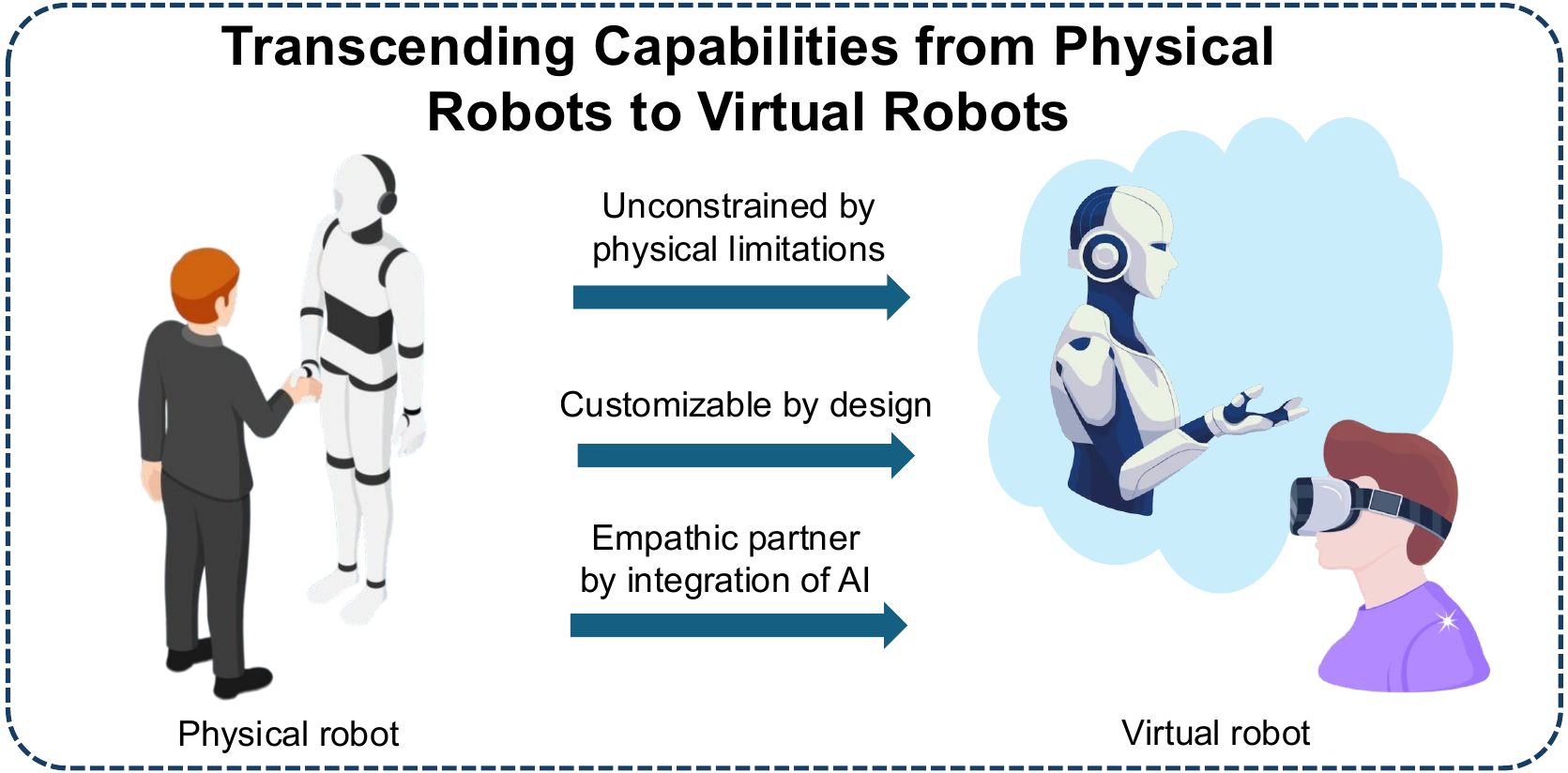}
    \caption{XR virtual agentic robots surpass physical capabilities.}
    \label{phy}
\end{figure}

\subsection{Surpassing Physical Robot Capabilities Through XR and AI Integration}
Another revolutionary implication of XR-based virtual robots is envisioned to be their ability to transcend the limitations of physical embodiments \cite{lee2006physically,wainer2006role}. Unconstrained by friction, mass, structural rigidity, or mechanical complexity, virtual agents are not bound to a single physical body or configuration \cite{zhang2021supporting}; they can transform roles instantaneously, adapt behaviors dynamically, and evolve across time and contexts. In VR environments, while the entire surroundings can be constructed virtually with immersion, designers are allowed either to mirror existing physical robots in appearance and shapes or to create entirely novel embodiments tailored to specific interaction goals \cite{kilteni2012sense,han2023crossing}. Locomotion, morphology, and other functions are therefore determined by the design intent customized for concrete needs rather than hardware feasibility. In AR and MR, similar principles apply: virtual counterpart robots or purely virtual agentic avatars can be anchored into the physical environment \cite{wolf2022exploring,nimcharoen2018me,genay2021being,zhang2022site}, moving and interacting according to the programmability that do not need to comply with the same constraints as physical robots. \\

These agents can seamlessly shift roles -- from tutor to collaborator to supervisor to others -- without the need for physical redesign, complex reconfiguration, or new hardware deployments. Through integration with LLMs or multimodal VLMs, they acquire the capacity to bear multi-turn dialogues, reason over task contexts, and anticipate user needs at a high level of abstraction \cite{konenkov2024vr}. Such abilities position XR-native robots as empathic mediators unconstrained by physical hardware: they not only deliver instructions and guidance, but also co-construct knowledge within specific contexts while sidestepping the limitations imposed by physical embodiments \cite{ghafurian2021improving,lisetti2013can,parmar2022designing}. In this sense, virtual robots act as adaptive, empathic partners that extend and complement human capabilities (Figure ~\ref{phy}).

\begin{figure}[!t]
    \centering
    \includegraphics[width=\textwidth]{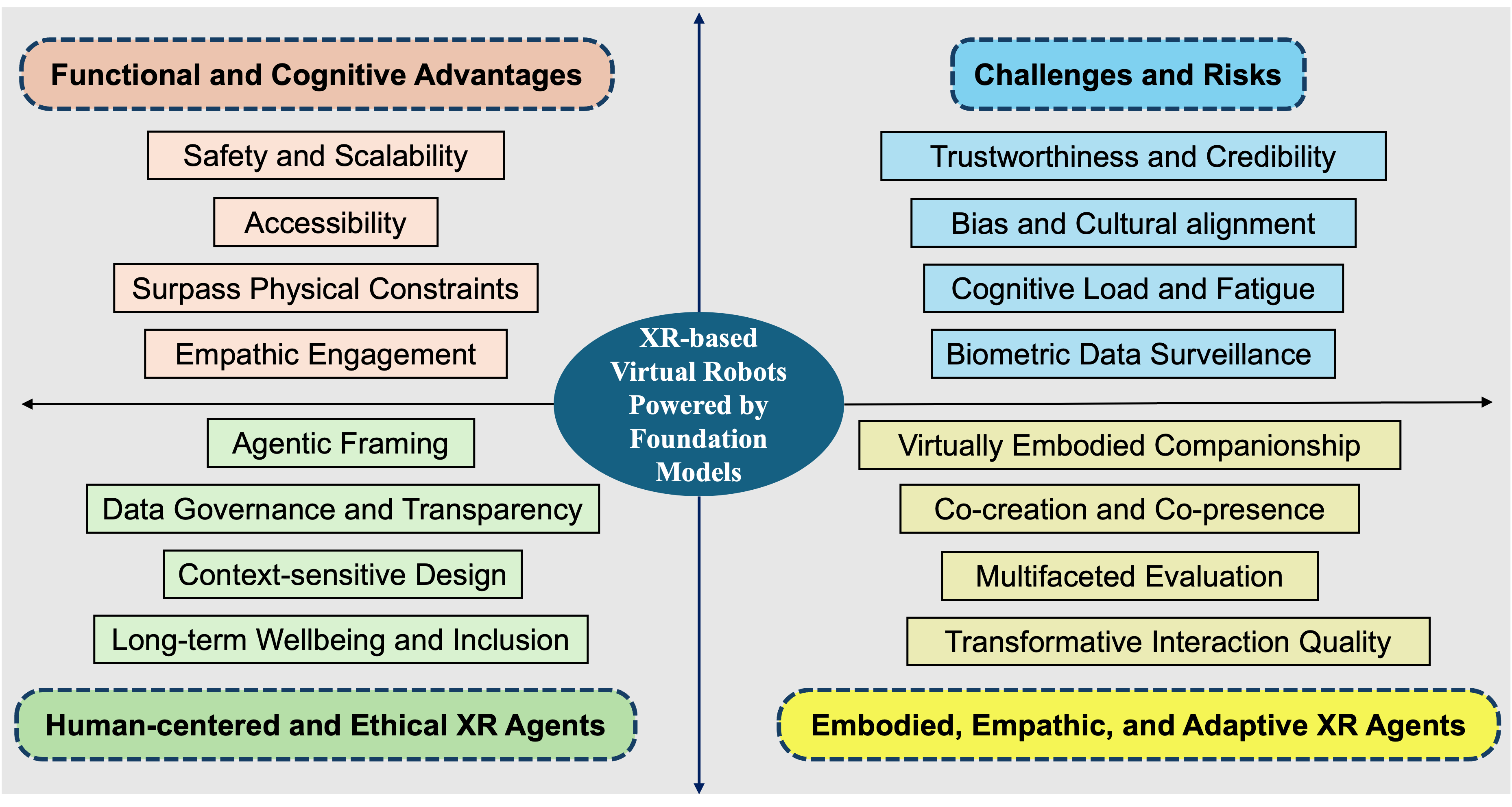}
    \caption{A new form of XR-based virtual robots powered by FMs. The upper left highlights functional and cognitive advantages, while the upper right summarizes key challenges. The lower bands illustrate two complementary design trajectories: human-centered and ethical XR agents (left) and embodied, empathic, and adaptive XR agents (right).}
    \label{pro_con}
\end{figure}

\section{Discussion}
To reframe the next generation of effective interactions between humans and virtual robots with XR, large FMs are expected to be employed for endowing virtual agents with situational awareness. This enables new forms of physical–virtual interaction, but also introduces challenges and risks, as illustrated in Figure~\ref{pro_con}.

\subsection*{Functional and Cognitive Advantages}
The integration of XR with AI-powered virtual robots offers advantages that are not merely incremental extensions of traditional HRI, but qualitatively new affordances. Functionally, safety and scalability remain the most immediate benefits: hazardous, rare, or high-stakes events can be rehearsed repeatedly without endangering humans or equipment. For example, industrial collaborative welding, chemical processing, and other risk-aware scenarios involving robots can be practiced in VR with virtual counterparts before an operator is ever exposed to a physical system.\\

On the cognitive side, XR-native agents powered by LLMs and VLMs can adapt to external input such as a user’s affective state, shifting from a neutral, task-focused presence to an empathic, supportive partner responsive to subtle cues. Similar mechanisms apply in social, therapeutic, and educational contexts, where virtual robots can dynamically adjust conversational depth, interaction tempo, or emotional tone based on user behavior. Moreover, the ability to continuously modify and personalize virtual robots makes XR an attractive environment for agile prototyping and co-design, where complex multi-agent interactions can be iteratively explored in simulation before committing to physical hardware or interface designs.

\subsection*{Virtual Embodiment and User Presence}
Although virtual robots lack physical mass, they are not disembodied abstractions. In XR, virtual embodiment is achieved through consistent visual appearance, coherent spatial behavior, and responsiveness that aligns with human expectations for social interaction \cite{zhang2023virtuality,wallace2024imitation}. Studies of MR hazard visualization and robot collaboration show that gaze synchrony, joint attention cues, and predictable proximity dynamics can substantially enhance users’ sense of safety, control, and shared situational awareness. Work on avatar responsiveness and motion timing similarly suggests that even relatively simple agents can evoke strong impressions of co-presence when they react promptly and contingently to user actions \cite{suzuki2023xr}.\\

However, this form of embodiment remains partial. The lack of rich haptic and tactile feedback limits the kinds of tasks that can be trained purely in XR, particularly those requiring fine manipulation, force sensing, or shared physical contact. Emerging haptic overlays and pseudo-physical resistance may narrow this gap, but they also introduce new layers of technical complexity and potential sensory mismatch. Designing virtual embodiment thus becomes a balancing act: providing sufficient realism to support user trust, learning, and empathy, without promising physical capabilities that the virtual environment cannot reliably deliver.

\subsection*{Potential Challenges and Risks}
The same characteristics that make XR-based virtual robots powered by large FMs compelling also introduce significant risks. A core concern is trustworthiness and credibility, closely mirroring broader issues in contemporary AI. LLM-driven agents can produce fluent, confident explanations even when their underlying reasoning is flawed or when they inherit biases from their training data. Cultural alignment presents another latent risk: models trained on skewed datasets may inadvertently reproduce stereotypes, misinterpret culturally specific gestures, or privilege certain communication styles over others. When these biases are embodied in seemingly empathic agents, they may go unnoticed yet subtly shape users’ expectations, self-perceptions, and sense of belonging. In XR, where users are immersed and may experience agents as co-present social partners, this illusion of competence can be especially potent. In safety-critical or therapeutic contexts, an erroneous suggestion delivered by a convincing virtual agent could have serious real-world consequences.\\

XR also amplifies familiar challenges around cognitive load and fatigue, similar to those documented in traditional XR applications. While immersive environments can enhance focus and memory, prolonged use has been associated with motion sickness, eye strain, and attentional exhaustion \cite{lopes2024computer}. Vulnerable populations—such as children, older adults, or individuals with certain neurological conditions—may be particularly sensitive to these effects. Furthermore, many XR systems rely on extensive biometric sensing, including gaze tracking, facial expression analysis, and voice-based emotion inference. Without careful design, governance, and transparency, these capabilities raise concerns about privacy, consent, and accessibility. The practicality and fairness of deploying XR-powered virtual robots to target groups therefore remains an open question, especially in domains such as healthcare and pediatric care where trust and vulnerability are heightened.

\subsection*{Toward Human-Centered and Ethical XR Agents}
The emergence of XR-based virtual robots powered by FMs introduces a new form of agency, autonomy, and perceived authenticity in HCI. As these agents begin to mirror human social cues -- maintaining eye contact, modulating voice, mirroring posture -- they encourage users to treat them as intentional partners rather than mere tools. This blurring of boundaries raises practical questions for designers and practitioners: How should such agents be framed and introduced to users? How can we ensure that the data used for pre-training and adaptation is curated, documented, and governed in ethically robust and trustworthy ways?\\

Different application domains may require distinct design principles. In therapy or education, for instance, some argue that strongly anthropomorphic agents risk fostering unhealthy attachment or unrealistic expectations, and therefore advocate for stylized, clearly artificial embodiments. Others argue that trust and participation—particularly for children, anxious patients, or socially isolated individuals—may depend on agents who feel warm, relatable, and socially competent. These tensions suggest that “one-size-fits-all” guidelines are unlikely to be sufficient. Instead, ethical frameworks should be context-sensitive, specifying appropriate levels of anthropomorphism, transparency, disclosure, and autonomy for each use case.\\

Beyond individual interactions, there is a broader ethical question surrounding virtual robots driven by versatile FMs capable of sensing, communicating, and providing companionship. The datasets and interaction logs that fuel these capabilities could be repurposed for surveillance, targeted persuasion, or other forms of misuse, particularly with respect to privacy and data protection. Addressing these concerns requires input from AI ethics, law and policy, developmental psychology, accessibility research, and design practice, among other fields. The goal is not only to prevent harm, but to ensure that XR-based agents are deployed in ways that are genuinely human-centered, supporting user agency, dignity, and long-term wellbeing rather than merely showcasing technical feasibility.

\subsection*{Toward Embodied, Empathic, and Adaptive XR Agents}
These concerns and opportunities point toward a reconceptualization of robots—from purely physical tools to relational, cognitively and emotionally engaged virtual partners. In XR, virtual robots do not primarily manipulate the physical world; instead, they shape users’ attention, scaffold their understanding, and participate in the co-construction of tasks and meanings. Their value lies in adaptive dialogue, emotional resonance, and the ability to inhabit shared virtual spaces in which roles, perspectives, and interaction scripts can be fluid and negotiable.\\

Designing for this paradigm demands new forms of co-creation and evaluation. On the design side, XR–HRI systems should be developed through interdisciplinary collaboration that includes not only HCI and AI researchers, but also domain experts such as clinicians, educators, and accessibility specialists, together with end users themselves. On the evaluation side, traditional performance metrics such as task completion time or error rate must be complemented by measures of psychological safety, perceived empathy, inclusion, and long-term behavioral and attitudinal change. Longitudinal studies are particularly needed to understand how repeated interaction with virtual agents affects user trust, skill retention, and social expectations over months or years.\\

Reframing virtual robots as relational entities rather than passive tools shifts the focus of innovation from mechanical capability to the quality of interaction. It requires asking not only what robots can do, but how they make people feel, what forms of understanding they enable, and whose needs they foreground or neglect. It is in answering these questions—empirically, ethically, and collaboratively—that the transformative potential of XR-based HRI is most likely to be realized.

\section{Future Research Agenda}
To ensure that XR-based virtual robots evolve in ethically responsible and socially beneficial ways, several research priorities can be pursued in the coming years. Building on the perspectives outlined in this paper, we highlight four potential research directions.

\subsection*{Evaluation Frameworks for Trust, Empathy, and Transferability}
Existing XR–HRI evaluations often foreground usability or task performance, but rarely capture the affective, relational, and longitudinal dimensions that are central to XR-native virtual robots. Future work should therefore develop multi-layered evaluation frameworks that address at least three questions: (i) how users calibrate trust in agents that may be communicatively fluent but epistemically unstable; (ii) how empathic relation and perceived psychological safety emerge and are maintained over time; and (iii) how learning and behavioral patterns acquired in virtual settings transfer to physical contexts. \\

Such frameworks will likely need to incorporate and coordinate behavioral, physiological, and self-report measures. Transferability studies are particularly underdeveloped in XR-HRI works: it remains unclear when and how competencies developed with XR-native virtual robots be generalized into interactions with physical robots or human teammates in real-world conditions. Addressing these gaps will require controlled virtuality-to-physicality experiments, longitudinal field deployments, and shared benchmarks that allow comparison across systems and domains.

\subsection*{Multi-Agent and Multi-User Interactions}
Real-world collaboration rarely occurs in dyads; it typically involves heterogeneous groups of humans and agents coordinating under uncertainty. Yet most XR–HRI prototypes still assume a single user and a homogeneous agent or agent group. Next-generation platforms should therefore support synchronous, parallel interaction among multiple users and multiple virtual robots, with roles and behaviors that adapt dynamically over time. This calls for mechanisms that enable flexible role allocation and reallocation between humans and agents, and for systematic research on how such roles are negotiated and transformed—how participants shift between leader, peer, tutor, and observer—and how these dynamics shape system accountability, perceived empathy, and inclusion. Understanding and deliberately designing for these emergent, multi-party interaction patterns will be crucial for safe, equitable, and effective XR-mediated teamwork.

\subsection*{Societal and Ethical Design}
As XR-based virtual robots might flourish at schools, clinics, workplaces, and homes, they will not only reflect but also participate in shaping social norms, expectations, and mutual relations. Future research must therefore move beyond ad hoc “ethics checklists” and toward empirically grounded, participatory approaches to socio-technical design. A first priority is the detection, characterization, and mitigation of bias in both reasoning processes and actual embodiments. This includes examining how cultural, gendered, racialized, or ability-related cues are encoded in appearance, voice, posture, and interaction style, and how these cues influence user trust, identification, and perceived legitimacy. \\

Second, the extensive biometric sensing often involved in XR -- eye tracking, facial expression analysis, posture and gesture monitoring, voice sentiment inference -- demands consent frameworks and data governance models that are legible to non-experts and co-developed with affected communities. Questions of data retention, secondary use, and the right to opt out of sensor-based personalization will be central, particularly in healthcare, education, and workplace surveillance contexts.

\subsection*{Mixed Embodiment and XR–Physical Integration}
Finally, the integration of XR-native agents with physical robots presents an important, yet comparatively underexplored frontier. XR can function as a cognitive and social twin of physical systems, providing a training and interaction layer that mirrors (or intentionally extends) the capabilities of real-world robots. Through shared autonomy and digital twin frameworks, users could engage with a virtual robot in XR -- developing shared routines, understanding capabilities, and calibrating trust -- before completely transitioning to interaction with its physical counterpart. \\

Research in this area must examine how consistency, continuity, and synchronization are maintained across modalities, including: When does divergence between virtual and physical behavior support learning and creativity; When does it undermine trust or induce negative awareness? How should haptic overlays, audio cues, and other multisensory feedback be orchestrated so that transitions between virtual and physical embodiments remain intelligible and safe? Addressing these questions will require joint advances in control, interface design, perception, and systematic evaluation, as well as careful consideration of where hybrid systems are most appropriate versus where purely virtual or purely physical approaches suffice.

\section{Conclusion and Outlook}
Virtual robots in XR mark a significant evolution in the broader spectrum of HRI. More than simulations or visualization tools, these agents can act as collaborative partners -- capable of learning from users, adapting to context, and engaging across cognitive, emotional, and spatial dimensions -- by the state-of-the-art large FMs. They offer safer environments for training and exploration, richer and more flexible educational and therapeutic experiences, and new spaces for participatory design and experimentation that are not constrained by the limits of physical hardware. \\

In this paper, we have argued that virtual robots in XR should be understood not only as technical artifacts, but as relational and cultural ones. They mediate how people learn, work, and relate to one another, and thus must be designed and evaluated with attention to empathy, inclusion, and long-term wellbeing. We call on researchers, designers, practitioners, and policymakers to treat XR-native virtual robots as part of a broader socio-technical ecosystem: one that spans AI, HRI, XR, ethics, accessibility, and domain expertise. By approaching these systems as sites of interdisciplinary co-creation -- rather than as isolated engineering projects -- we can better ensure that their evolution is aligned with human values and diverse forms of flourishing. The future of virtual robots, like our own, will be shaped by the commitments we adopt and the infrastructures we build in the present. \\

At the same time, this potential is inseparable from responsibility. Without appropriate design and governance, XR-based virtual robots with large FMs may inadvertently reproduce existing inequities, foster less transparency, or blur the boundaries in certain aspects. As XR technologies mature and FMs grow more capable and autonomous, we are approaching a decisive juncture: whether virtual robots will amplify human cognition, compassion, and creativity, or raise bias, erode agency, and normalize intrusive sensing will depend on choices made now.

\section*{Declarations}

\subsection*{Acknowledgements}
We acknowledge that authors made limited use of generative AI tools for language polishing only.

\subsection*{Authors contribution}
Conceptualization: Y.Z.; Materials preparation: Y.Z. and Y.M.; Writing: Y.Z.; Supervision: D.K..

\subsection*{Conflicts of interest}
The authors declare no conflicts of interest.

\subsection*{Ethical approval}
Not applicable.

\subsection*{Consent to participate}
Not applicable.

\subsection*{Consent for publication}
Not applicable.

\subsection*{Availability of data and materials}
Not applicable.

\subsection*{Funding}
None.

\subsubsection*{Copyright}
\copyright The Author(s) 2025.

\bibliography{template}

@inproceedings{dianatfar2024virtual,
  title={Virtual reality-based safety training in human-robot collaboration scenario: User experiences testing},
  author={Dianatfar, Morteza and P{\"o}ys{\"a}ri, Saku and Latokartano, Jyrki and Siltala, Niko and Lanz, Minna},
  booktitle={AIP Conference Proceedings},
  volume={2989},
  number={1},
  pages={020002},
  year={2024},
  organization={AIP Publishing LLC}
}

@inproceedings{zhang2023large,
  title={Large language models as zero-shot human models for human-robot interaction},
  author={Zhang, Bowen and Soh, Harold},
  booktitle={2023 IEEE/RSJ international conference on intelligent robots and systems (IROS)},
  pages={7961--7968},
  year={2023},
  organization={IEEE}
}

@article{wang2024large,
  title={Large language models for robotics: Opportunities, challenges, and perspectives},
  author={Wang, Jiaqi and Shi, Enze and Hu, Huawen and Ma, Chong and Liu, Yiheng and Wang, Xuhui and Yao, Yincheng and Liu, Xuan and Ge, Bao and Zhang, Shu},
  journal={Journal of Automation and Intelligence},
  year={2024},
  publisher={Elsevier}
}

@article{san2025mixed,
  title={Mixed reality representation of hazard zones while collaborating with a robot: sense of control over own safety},
  author={San Martin, Ane and Kildal, Johan and Lazkano, Elena},
  journal={Virtual Reality},
  volume={29},
  number={1},
  pages={1--20},
  year={2025},
  publisher={Springer}
}

@article{konenkov2024vr,
  title={VR-GPT: Visual Language Model for Intelligent Virtual Reality Applications},
  author={Konenkov, M and Lykov, A and Trinitatova, D and Tsetserukou, D},
  journal={arXiv preprint arXiv:2405.11537},
  year={2024}
}

@inproceedings{buldu2025cuify,
  title={Cuify the xr: An open-source package to embed llm-powered conversational agents in xr},
  author={Buldu, Kadir Burak and {\"O}zdel, S{\"u}leyman and Lau, Ka Hei Carrie and Wang, Mengdi and Saad, Daniel and Sch{\"o}nborn, Sofie and Boch, Auxane and Kasneci, Enkelejda and Bozkir, Efe},
  booktitle={2025 IEEE International Conference on Artificial Intelligence and eXtended and Virtual Reality (AIxVR)},
  pages={192--197},
  year={2025},
  organization={IEEE}
}

@article{xie2024new,
  title={A new XR-based human-robot collaboration assembly system based on industrial metaverse},
  author={Xie, Jiacheng and Liu, Yali and Wang, Xuewen and Fang, Shukai and Liu, Shuguang},
  journal={Journal of Manufacturing Systems},
  volume={74},
  pages={949--964},
  year={2024},
  publisher={Elsevier}
}

@article{afzal2025next,
  title={Next Generation XR Systems—Large Language Models Meet Augmented and Virtual Reality},
  author={Afzal, Muhamamd Zeshan and Ali, SK Aziz and Stricker, Didier and others},
  journal={IEEE Computer Graphics and Applications},
  volume={43},
  number={1},
  pages={43--49},
  year={2025},
  doi={10.1109/MCG.2025.3548554},
  publisher={IEEE}
}

@inproceedings{li2024vision,
  title={Vision-Language Foundation Models as Effective Robot Imitators},
  author={Li, Xinghang and Liu, Minghuan and Zhang, Hanbo and Yu, Cunjun and Xu, Jie and Wu, Hongtao and Cheang, Chilam and Jing, Ya and Zhang, Weinan and Liu, Huaping and others},
  booktitle={ICLR},
  year={2024}
}

@inproceedings{suzuki2023xr,
  title={XR and AI: AI-Enabled Virtual, Augmented, and Mixed Reality},
  author={Suzuki, Ryo and Gonzalez-Franco, Mar and Sra, Misha and Lindlbauer, David},
  booktitle={Adjunct Proceedings of the 36th Annual ACM Symposium on User Interface Software and Technology (UIST)},
  pages={1--3},
  year={2023},
  organization={ACM},
  doi={10.1145/3586182.3617432}
}

@article{wang2025systematic,
  title={A Systematic Review of XR-Enabled Remote Human-Robot Interaction Systems},
  author={Wang, Xian and Shen, Luyao and Lee, Lik-Hang},
  journal={ACM Computing Surveys},
  volume={57},
  number={11},
  pages={1--37},
  year={2025},
  publisher={ACM New York, NY}
}

@article{pan2024integrating,
  title={Integrating extended reality and robotics in construction: A critical review},
  author={Pan, Mi and Wong, Mun On and Lam, Chi Chiu and Pan, Wei},
  journal={Advanced Engineering Informatics},
  volume={62},
  pages={102795},
  year={2024},
  publisher={Elsevier}
}

@article{lopes2024computer,
  title={Computer Vision in Augmented, Virtual, Mixed and Extended Reality environments—A bibliometric review},
  author={Lopes, J{\'u}lio Castro and Lopes, Rui Pedro},
  journal={Visual Informatics},
  volume={8},
  number={4},
  pages={13--22},
  year={2024},
  publisher={Elsevier}
}

@inproceedings{karpichev2024extended,
  title={Extended reality for enhanced human-robot collaboration: a human-in-the-loop approach},
  author={Karpichev, Yehor and Charter, Todd and Hong, Jayden and Enayati, Amir M Soufi and Honari, Homayoun and Tamizi, Mehran Ghafarian and Najjaran, Homayoun},
  booktitle={2024 33rd IEEE International Conference on Robot and Human Interactive Communication (ROMAN)},
  pages={1991--1998},
  year={2024},
  organization={IEEE}
}

@inproceedings{choi2022xr,
  title={An xr-based approach to safe human-robot collaboration},
  author={Choi, Sung Ho and Park, Kyeong-Beom and Roh, Dong Hyeon and Lee, Jae Yeol and Ghasemi, Yalda and Jeong, Heejin},
  booktitle={2022 IEEE Conference on Virtual Reality and 3D User Interfaces Abstracts and Workshops (VRW)},
  pages={481--482},
  year={2022},
  organization={IEEE}
}

@ARTICLE{Lakhnati2024exploring,
AUTHOR={Lakhnati, Younes  and Pascher, Max  and Gerken, Jens },        
TITLE={Exploring a GPT-based large language model for variable autonomy in a VR-based human-robot teaming simulation},        
JOURNAL={Frontiers in Robotics and AI},      
VOLUME={Volume 11 - 2024},
YEAR={2024},
URL={https://www.frontiersin.org/journals/robotics-and-ai/articles/10.3389/frobt.2024.1347538},
DOI={10.3389/frobt.2024.1347538},
ISSN={2296-9144},
}

@article{ZHANG2023100131,
title = {Large language models for human–robot interaction: A review},
journal = {Biomimetic Intelligence and Robotics},
volume = {3},
number = {4},
pages = {100131},
year = {2023},
issn = {2667-3797},
doi = {https://doi.org/10.1016/j.birob.2023.100131},
url = {https://www.sciencedirect.com/science/article/pii/S2667379723000451},
author = {Ceng Zhang and Junxin Chen and Jiatong Li and Yanhong Peng and Zebing Mao},
}

@INPROCEEDINGS{Murnane2021_virtual,
  author={Murnane, Mark and Higgins, Padraig and Saraf, Monali and Ferraro, Francis and Matuszek, Cynthia and Engel, Don},
  booktitle={2021 IEEE Conference on Virtual Reality and 3D User Interfaces Abstracts and Workshops (VRW)}, 
  title={A Simulator for Human-Robot Interaction in Virtual Reality}, 
  year={2021},
  volume={},
  number={},
  pages={470-471},
  doi={10.1109/VRW52623.2021.00117}}

@article{Higgins2021_virtual,
place = {Country unknown/Code not available}, 
title = {Towards Making Virtual Human-Robot Interaction a Reality}, 
url = {https://par.nsf.gov/biblio/10216761}, 
journal = {Proc. of the 3rd International Workshop on Virtual, Augmented, and Mixed-Reality for Human-Robot Interactions (VAM-HRI)}, 
author = {Higgins, Padraig and Kebe, Gaoussou Youssouf and Berlier, Adam and Darvish, Kasra and Engel, Don and Ferraro, Francis and Matuszek, Cynthia}, 
year={2021},
}

@article{wang2011handshake,
  title={Handshake: Realistic human-robot interaction in haptic enhanced virtual reality},
  author={Wang, Zheng and Giannopoulos, Elias and Slater, Mel and Peer, Angelika},
  journal={Presence},
  volume={20},
  number={4},
  pages={371--392},
  year={2011},
  publisher={MIT Press}
}

@inproceedings{villani2018use,
  title={Use of virtual reality for the evaluation of human-robot interaction systems in complex scenarios},
  author={Villani, Valeria and Capelli, Beatrice and Sabattini, Lorenzo},
  booktitle={2018 27th IEEE international symposium on robot and human interactive communication (RO-MAN)},
  pages={422--427},
  year={2018},
  organization={IEEE}
}

@article{walker2023virtual,
  title={Virtual, augmented, and mixed reality for human-robot interaction: A survey and virtual design element taxonomy},
  author={Walker, Michael and Phung, Thao and Chakraborti, Tathagata and Williams, Tom and Szafir, Daniel},
  journal={ACM Transactions on Human-Robot Interaction},
  volume={12},
  number={4},
  pages={1--39},
  year={2023},
  publisher={ACM New York, NY}
}

@inproceedings{chacko2019augmented,
  title={An augmented reality interface for human-robot interaction in unconstrained environments},
  author={Chacko, Sonia Mary and Kapila, Vikram},
  booktitle={2019 IEEE/RSJ International Conference on Intelligent Robots and Systems (IROS)},
  pages={3222--3228},
  year={2019},
  organization={IEEE}
}

@inproceedings{sabbella2023virtual,
  title={Virtual Reality Applications for Enhancing Human-Robot Interaction: A Gesture Recognition Perspective},
  author={Sabbella, Sandeep Reddy and Kaszuba, Sara and Leotta, Francesco and Nardi, Daniele},
  booktitle={Proceedings of the 23rd ACM International Conference on Intelligent Virtual Agents},
  pages={1--4},
  year={2023}
}

@inproceedings{arntz2024enhancing,
  title={Enhancing human-robot interaction research by using a virtual reality lab approach},
  author={Arntz, Alexander and Helgert, Andr{\'e} and Stra{\ss}mann, Carolin and Eimler, Sabrina C},
  booktitle={2024 IEEE International Conference on Artificial Intelligence and eXtended and Virtual Reality (AIxVR)},
  pages={340--344},
  year={2024},
  organization={IEEE}
}

@article{bustamante2021armsym,
  title={ArmSym: A virtual human--robot interaction laboratory for assistive robotics},
  author={Bustamante, Samuel and Peters, Jan and Sch{\"o}lkopf, Bernhard and Grosse-Wentrup, Moritz and Jayaram, Vinay},
  journal={IEEE Transactions on Human-Machine Systems},
  volume={51},
  number={6},
  pages={568--577},
  year={2021},
  publisher={IEEE}
}

@article{malik2021digital,
  title={Digital twins for collaborative robots: A case study in human-robot interaction},
  author={Malik, Ali Ahmad and Brem, Alexander},
  journal={Robotics and Computer-Integrated Manufacturing},
  volume={68},
  pages={102092},
  year={2021},
  publisher={Elsevier}
}

@inproceedings{fratczak2019understanding,
  title={Understanding human behaviour in industrial human-robot interaction by means of virtual reality},
  author={Fratczak, Piotr and Goh, Yee Mey and Kinnell, Peter and Soltoggio, Andrea and Justham, Laura},
  booktitle={Proceedings of the Halfway to the Future Symposium 2019},
  pages={1--7},
  year={2019}
}

@InProceedings{Duguleana2011eva,
author="Duguleana, Mihai and Barbuceanu, Florin Grigorie and Mogan, Gheorghe",
editor="Shumaker, Randall",
title="Evaluating Human-Robot Interaction during a Manipulation Experiment Conducted in Immersive Virtual Reality",
booktitle="Virtual and Mixed Reality - New Trends",
year="2011",
publisher="Springer Berlin Heidelberg",
address="Berlin, Heidelberg",
pages="164--173",
}

@inproceedings{bolano2019transparent_virtual,
  title={Transparent robot behavior using augmented reality in close human-robot interaction},
  author={Bolano, Gabriele and Juelg, Christian and Roennau, Arne and Dillmann, Ruediger},
  booktitle={2019 28th IEEE International Conference on Robot and Human Interactive Communication (RO-MAN)},
  pages={1--7},
  year={2019},
  organization={IEEE}
}

@inproceedings{green2007human,
  title={Human robot collaboration: An augmented reality approach—a literature review and analysis},
  author={Green, Scott A and Billinghurst, Mark and Chen, XiaoQi and Chase, J Geoffrey},
  booktitle={International design engineering technical conferences and computers and information in engineering conference},
  volume={48051},
  pages={117--126},
  year={2007}
}

@article{fang2014novel,
  title={A novel augmented reality-based interface for robot path planning},
  author={Fang, HC and Ong, SK and Nee, AYC},
  journal={International Journal on Interactive Design and Manufacturing (IJIDeM)},
  volume={8},
  number={1},
  pages={33--42},
  year={2014},
  publisher={Springer}
}

@INPROCEEDINGS{Milgram1993,
  author={Milgram, P. and Zhai, S. and Drascic, D. and Grodski, J.},
  booktitle={Proceedings of 1993 IEEE/RSJ International Conference on Intelligent Robots and Systems (IROS '93)}, 
  title={Applications of augmented reality for human-robot communication}, 
  year={1993},
  volume={3},
  number={},
  pages={1467-1472 vol.3},
  doi={10.1109/IROS.1993.583833}}

@inproceedings{walker2019robot,
  title={Robot teleoperation with augmented reality virtual surrogates},
  author={Walker, Michael E and Hedayati, Hooman and Szafir, Daniel},
  booktitle={2019 14th ACM/IEEE International Conference on Human-Robot Interaction (HRI)},
  pages={202--210},
  year={2019},
  organization={IEEE}
}

@inproceedings{suzuki2022augmented,
  title={Augmented reality and robotics: A survey and taxonomy for ar-enhanced human-robot interaction and robotic interfaces},
  author={Suzuki, Ryo and Karim, Adnan and Xia, Tian and Hedayati, Hooman and Marquardt, Nicolai},
  booktitle={Proceedings of the 2022 CHI Conference on Human Factors in Computing Systems},
  pages={1--33},
  year={2022}
}

@article{guhl2018enabling,
  title={Enabling human-robot-interaction via virtual and augmented reality in distributed control systems},
  author={Guhl, Jan and H{\"u}gle, Johannes and Kr{\"u}ger, J{\"o}rg},
  journal={Procedia CIRP},
  volume={76},
  pages={167--170},
  year={2018},
  publisher={Elsevier}
}

@inproceedings{wang2015intelligent,
  title={Intelligent agents for virtual simulation of human-robot interaction},
  author={Wang, Ning and Pynadath, David V and Unnikrishnan, KV and Shankar, Santosh and Merchant, Chirag},
  booktitle={International Conference on Virtual, Augmented and Mixed Reality},
  pages={228--239},
  year={2015},
  organization={Springer}
}

@article{groechel2022tool,
  title={A tool for organizing key characteristics of virtual, augmented, and mixed reality for human--robot interaction systems: Synthesizing vam-hri trends and takeaways},
  author={Groechel, Thomas R and Walker, Michael E and Chang, Christine T and Rosen, Eric and Forde, Jessica Zosa},
  journal={IEEE Robotics \& Automation Magazine},
  volume={29},
  number={1},
  pages={35--44},
  year={2022},
  publisher={IEEE}
}

@inproceedings{szafir2019mediating,
  title={Mediating human-robot interactions with virtual, augmented, and mixed reality},
  author={Szafir, Daniel},
  booktitle={International Conference on Human-Computer Interaction},
  pages={124--149},
  year={2019},
  organization={Springer}
}

@article{kousi2019enabling,
  title={Enabling human robot interaction in flexible robotic assembly lines: An augmented reality based software suite},
  author={Kousi, Niki and Stoubos, Christos and Gkournelos, Christos and Michalos, George and Makris, Sotiris},
  journal={Procedia CIRP},
  volume={81},
  pages={1429--1434},
  year={2019},
  publisher={Elsevier}
}

@INPROCEEDINGS{Maly2016,
  author={Malý, Ivo and Sedláček, David and Leitão, Paulo},
  booktitle={2016 IEEE 14th International Conference on Industrial Informatics (INDIN)}, 
  title={Augmented reality experiments with industrial robot in industry 4.0 environment}, 
  year={2016},
  volume={},
  number={},
  pages={176-181},
  doi={10.1109/INDIN.2016.7819154}}

@inproceedings{walker2018communicating,
  title={Communicating robot motion intent with augmented reality},
  author={Walker, Michael and Hedayati, Hooman and Lee, Jennifer and Szafir, Daniel},
  booktitle={Proceedings of the 2018 ACM/IEEE International Conference on Human-Robot Interaction},
  pages={316--324},
  year={2018}
}

@inproceedings{manring2019augmented,
  title={Augmented reality for interactive robot control},
  author={Manring, Levi and Pederson, John and Potts, Dillon and Boardman, Beth and Mascarenas, David and Harden, Troy and Cattaneo, Alessandro},
  booktitle={Special Topics in Structural Dynamics \& Experimental Techniques, Volume 5: Proceedings of the 37th IMAC, A Conference and Exposition on Structural Dynamics 2019},
  pages={11--18},
  year={2019},
  organization={Springer}
}

@article{wang2024robotic,
  title={A robotic teleoperation system enhanced by augmented reality for natural human--robot interaction},
  author={Wang, Xingchao and Guo, Shuqi and Xu, Zijian and Zhang, Zheyuan and Sun, Zhenglong and Xu, Yangsheng},
  journal={Cyborg and Bionic Systems},
  volume={5},
  pages={0098},
  year={2024},
  publisher={AAAS}
}

@article{fang2022brain,
  title={Brain--computer interface integrated with augmented reality for human--robot interaction},
  author={Fang, Bin and Ding, Wenlong and Sun, Fuchun and Shan, Jianhua and Wang, Xiaojia and Wang, Chengyin and Zhang, Xinyu},
  journal={IEEE Transactions on Cognitive and Developmental Systems},
  volume={15},
  number={4},
  pages={1702--1711},
  year={2022},
  publisher={IEEE}
}

@inproceedings{bischoff2004perspectives,
  title={Perspectives on augmented reality based human-robot interaction with industrial robots},
  author={Bischoff, Rainer and Kazi, Arif},
  booktitle={2004 IEEE/RSJ International Conference on Intelligent Robots and Systems (IROS)(IEEE Cat. No. 04CH37566)},
  volume={4},
  pages={3226--3231},
  year={2004},
  organization={IEEE}
}

@article{hernandez2020increasing_virtual,
  title={Increasing robot autonomy via motion planning and an augmented reality interface},
  author={Hern{\'a}ndez, Juan David and Sobti, Shlok and Sciola, Anthony and Moll, Mark and Kavraki, Lydia E},
  journal={IEEE Robotics and Automation Letters},
  volume={5},
  number={2},
  pages={1017--1023},
  year={2020},
  publisher={IEEE}
}

@inproceedings{qiu2020human,
  title={Human-robot interaction in a shared augmented reality workspace},
  author={Qiu, Shuwen and Liu, Hangxin and Zhang, Zeyu and Zhu, Yixin and Zhu, Song-Chun},
  booktitle={2020 IEEE/RSJ International Conference on Intelligent Robots and Systems (IROS)},
  pages={11413--11418},
  year={2020},
  organization={IEEE}
}

@inproceedings{hedayati2018improving,
  title={Improving collocated robot teleoperation with augmented reality},
  author={Hedayati, Hooman and Walker, Michael and Szafir, Daniel},
  booktitle={Proceedings of the 2018 ACM/IEEE International Conference on Human-Robot Interaction},
  pages={78--86},
  year={2018}
}

@inproceedings{wang2019exploring,
  title={Exploring virtual agents for augmented reality},
  author={Wang, Isaac and Smith, Jesse and Ruiz, Jaime},
  booktitle={Proceedings of the 2019 CHI Conference on Human Factors in Computing Systems},
  pages={1--12},
  year={2019}
}

@INPROCEEDINGS{Arntz2020_virtual,
  author={Arntz, Alexander and Eimler, Sabrina C. and Hoppe, H. Ulrich},
  booktitle={2020 IEEE International Conference on Artificial Intelligence and Virtual Reality (AIVR)}, 
  title={“The Robot-Arm Talks Back to Me” - Human Perception of Augmented Human-Robot Collaboration in Virtual Reality}, 
  year={2020},
  volume={},
  number={},
  pages={307-312},
  doi={10.1109/AIVR50618.2020.00062}}

@article{yamamoto2012augmented,
  title={Augmented reality and haptic interfaces for robot-assisted surgery},
  author={Yamamoto, Tomonori and Abolhassani, Niki and Jung, Sung and Okamura, Allison M and Judkins, Timothy N},
  journal={The International Journal of Medical Robotics and Computer Assisted Surgery},
  volume={8},
  number={1},
  pages={45--56},
  year={2012},
  publisher={Wiley Online Library}
}

@article{michalos2016augmented,
  title={Augmented reality (AR) applications for supporting human-robot interactive cooperation},
  author={Michalos, George and Karagiannis, Panagiotis and Makris, Sotiris and Tok{\c{c}}alar, {\"O}nder and Chryssolouris, George},
  journal={Procedia CIRP},
  volume={41},
  pages={370--375},
  year={2016},
  publisher={Elsevier}
}

@article{ostanin2019interactive,
  title={Interactive robots control using mixed reality},
  author={Ostanin, Mikhail and Yagfarov, Rauf and Klimchik, Alexandr},
  journal={IFAC-PapersOnLine},
  volume={52},
  number={13},
  pages={695--700},
  year={2019},
  publisher={Elsevier}
}

@article{Ostanin02122021,
author = {M. Ostanin and R. Yagfarov and D. Devitt and A. Akhmetzyanov and A. Klimchik},
title = {Multi robots interactive control using mixed reality},
journal = {International Journal of Production Research},
volume = {59},
number = {23},
pages = {7126--7138},
year = {2021},
publisher = {Taylor \& Francis},
doi = {10.1080/00207543.2020.1834640},
}

@article{choi2022integrated,
  title={An integrated mixed reality system for safety-aware human-robot collaboration using deep learning and digital twin generation},
  author={Choi, Sung Ho and Park, Kyeong-Beom and Roh, Dong Hyeon and Lee, Jae Yeol and Mohammed, Mustafa and Ghasemi, Yalda and Jeong, Heejin},
  journal={Robotics and Computer-Integrated Manufacturing},
  volume={73},
  pages={102258},
  year={2022},
  publisher={Elsevier}
}

@article{Li2023assist,
title = {An AR-assisted Deep Reinforcement Learning-based approach towards mutual-cognitive safe human-robot interaction},
journal = {Robotics and Computer-Integrated Manufacturing},
volume = {80},
pages = {102471},
year = {2023},
issn = {0736-5845},
doi = {https://doi.org/10.1016/j.rcim.2022.102471},
url = {https://www.sciencedirect.com/science/article/pii/S0736584522001533},
author = {Chengxi Li and Pai Zheng and Yue Yin and Yat Ming Pang and Shengzeng Huo},
}

@inproceedings{peters2018towards_virtual,
  title={Towards the use of mixed reality for hri design via virtual robots},
  author={Peters, Christopher and Yang, Fangkai and Saikia, Himangshu and Li, Chengjie and Skantze, Gabriel},
  booktitle={1st International Workshop on Virtual, Augmented, and Mixed Reality for HRI (VAM-HRI), Cambridge, UK, March 23, 2020},
  year={2018}
}

@article{park2021hands,
  title={Hands-free human--robot interaction using multimodal gestures and deep learning in wearable mixed reality},
  author={Park, Kyeong-Beom and Choi, Sung Ho and Lee, Jae Yeol and Ghasemi, Yalda and Mohammed, Mustafa and Jeong, Heejin},
  journal={IEEE Access},
  volume={9},
  pages={55448--55464},
  year={2021},
  publisher={IEEE}
}

@article{lee2006physically,
  title={Are physically embodied social agents better than disembodied social agents?: The effects of physical embodiment, tactile interaction, and people's loneliness in human--robot interaction},
  author={Lee, Kwan Min and Jung, Younbo and Kim, Jaywoo and Kim, Sang Ryong},
  journal={International journal of human-computer studies},
  volume={64},
  number={10},
  pages={962--973},
  year={2006},
  publisher={Elsevier}
}

@inproceedings{wainer2006role,
  title={The role of physical embodiment in human-robot interaction},
  author={Wainer, Joshua and Feil-Seifer, David J and Shell, Dylan A and Mataric, Maja J},
  booktitle={ROMAN 2006-The 15th IEEE International Symposium on Robot and Human Interactive Communication},
  pages={117--122},
  year={2006},
  organization={IEEE}
}

@article{kilteni2012sense,
  title={The sense of embodiment in virtual reality},
  author={Kilteni, Konstantina and Groten, Raphaela and Slater, Mel},
  journal={Presence: Teleoperators and Virtual Environments},
  volume={21},
  number={4},
  pages={373--387},
  year={2012},
  publisher={MIT Press One Rogers Street, Cambridge, MA 02142-1209, USA journals-info~…}
}

@inproceedings{han2023crossing,
  title={Crossing reality: Comparing physical and virtual robot deixis},
  author={Han, Zhao and Zhu, Yifei and Phan, Albert and Garza, Fernando Sandoval and Castro, Amia and Williams, Tom},
  booktitle={Proceedings of the 2023 ACM/IEEE International Conference on Human-Robot Interaction},
  pages={152--161},
  year={2023}
}

@inproceedings{wolf2022exploring,
  title={Exploring presence, avatar embodiment, and body perception with a holographic augmented reality mirror},
  author={Wolf, Erik and Fiedler, Marie Luisa and D{\"o}llinger, Nina and Wienrich, Carolin and Latoschik, Marc Erich},
  booktitle={2022 IEEE conference on virtual reality and 3D user interfaces (VR)},
  pages={350--359},
  year={2022},
  organization={IEEE}
}

@inproceedings{nimcharoen2018me,
  title={Is that me?—Embodiment and body perception with an augmented reality mirror},
  author={Nimcharoen, Chontira and Zollmann, Stefanie and Collins, Jonny and Regenbrecht, Holger},
  booktitle={2018 IEEE International Symposium on Mixed and Augmented Reality Adjunct (ISMAR-Adjunct)},
  pages={158--163},
  year={2018},
  organization={IEEE}
}

@article{genay2021being,
  title={Being an avatar “for real”: a survey on virtual embodiment in augmented reality},
  author={Genay, Ad{\'e}la{\"\i}de and L{\'e}cuyer, Anatole and Hachet, Martin},
  journal={IEEE Transactions on Visualization and Computer Graphics},
  volume={28},
  number={12},
  pages={5071--5090},
  year={2021},
  publisher={IEEE}
}

@article{paiva2017empathy,
  title={Empathy in virtual agents and robots: A survey},
  author={Paiva, Ana and Leite, Iolanda and Boukricha, Hana and Wachsmuth, Ipke},
  journal={ACM Transactions on Interactive Intelligent Systems (TiiS)},
  volume={7},
  number={3},
  pages={1--40},
  year={2017},
  publisher={ACM New York, NY, USA}
}

@article{zhao2003toward,
  title={Toward a taxonomy of copresence},
  author={Zhao, Shanyang},
  journal={Presence},
  volume={12},
  number={5},
  pages={445--455},
  year={2003},
  publisher={MIT Press}
}

@inproceedings{weistroffer2014assessing,
  title={Assessing the acceptability of human-robot co-presence on assembly lines: A comparison between actual situations and their virtual reality counterparts},
  author={Weistroffer, Vincent and Paljic, Alexis and Fuchs, Philippe and Hugues, Olivier and Chodacki, Jean-Paul and Ligot, Pascal and Morais, Alexandre},
  booktitle={The 23rd IEEE International Symposium on Robot and Human Interactive Communication},
  pages={377--384},
  year={2014},
  organization={IEEE}
}

@article{szczurek2023multimodal,
  title={Multimodal multi-user mixed reality human--robot interface for remote operations in hazardous environments},
  author={Szczurek, Krzysztof Adam and Prades, Raul Marin and Matheson, Eloise and Rodriguez-Nogueira, Jose and Di Castro, Mario},
  journal={IEEE Access},
  volume={11},
  pages={17305--17333},
  year={2023},
  publisher={IEEE}
}

@article{simaan2015intelligent,
  title={Intelligent surgical robots with situational awareness},
  author={Simaan, Nabil and Taylor, Rusell H and Choset, Howie},
  journal={Mechanical Engineering},
  volume={137},
  number={09},
  pages={S3--S6},
  year={2015},
  publisher={American Society of Mechanical Engineers}
}

@article{roldan2017multi,
  title={Multi-robot interfaces and operator situational awareness: Study of the impact of immersion and prediction},
  author={Rold{\'a}n, Juan Jes{\'u}s and Pe{\~n}a-Tapia, Elena and Mart{\'\i}n-Barrio, Andr{\'e}s and Olivares-M{\'e}ndez, Miguel A and Del Cerro, Jaime and Barrientos, Antonio},
  journal={Sensors},
  volume={17},
  number={8},
  pages={1720},
  year={2017},
  publisher={MDPI}
}

@inproceedings{van2024puppeteer,
  title={Puppeteer your robot: Augmented reality leader-follower teleoperation},
  author={Van Haastregt, Jonne and Welle, Michael C and Zhang, Yuchong and Kragic, Danica},
  booktitle={2024 IEEE-RAS 23rd International Conference on Humanoid Robots (Humanoids)},
  pages={1019--1026},
  year={2024},
  organization={IEEE}
}

@inproceedings{zhang2025llm,
  title={LLM-Driven Augmented Reality Puppeteer: Controller-Free Voice-Commanded Robot Teleoperation},
  author={Zhang, Yuchong and Orthmann, Bastian and Welle, Michael C and Van Haastregt, Jonne and Kragic, Danica},
  booktitle={International Conference on Human-Computer Interaction},
  pages={97--112},
  year={2025},
  organization={Springer}
}

@article{zhang2025multimodal,
  title={Multimodal" Puppeteer": An Exploration of Robot Teleoperation Via Virtual Counterpart with LLM-Driven Voice and Gesture Interaction in Augmented Reality},
  author={Zhang, Yuchong and Orthmann, Bastian and Ji, Shichen and Welle, Michael and Van Haastregt, Jonne and Kragic, Danica},
  journal={arXiv preprint arXiv:2506.13189},
  year={2025}
}

@inproceedings{kim2024understanding,
  title={Understanding large-language model (llm)-powered human-robot interaction},
  author={Kim, Callie Y and Lee, Christine P and Mutlu, Bilge},
  booktitle={Proceedings of the 2024 ACM/IEEE international conference on human-robot interaction},
  pages={371--380},
  year={2024}
}

@article{kawaharazuka2024real,
  title={Real-world robot applications of foundation models: A review},
  author={Kawaharazuka, Kento and Matsushima, Tatsuya and Gambardella, Andrew and Guo, Jiaxian and Paxton, Chris and Zeng, Andy},
  journal={Advanced Robotics},
  volume={38},
  number={18},
  pages={1232--1254},
  year={2024},
  publisher={Taylor \& Francis}
}

@article{costa2022augmented,
  title={Augmented reality for human--robot collaboration and cooperation in industrial applications: A systematic literature review},
  author={Costa, Gabriel de Moura and Petry, Marcelo Roberto and Moreira, Ant{\'o}nio Paulo},
  journal={Sensors},
  volume={22},
  number={7},
  pages={2725},
  year={2022},
  publisher={MDPI}
}

@article{robins2005robotic,
  title={Robotic assistants in therapy and education of children with autism: can a small humanoid robot help encourage social interaction skills?},
  author={Robins, Ben and Dautenhahn, Kerstin and Boekhorst, R Te and Billard, Aude},
  journal={Universal access in the information society},
  volume={4},
  number={2},
  pages={105--120},
  year={2005},
  publisher={Springer}
}

@inproceedings{shamsuddin2012initial,
  title={Initial response of autistic children in human-robot interaction therapy with humanoid robot NAO},
  author={Shamsuddin, Syamimi and Yussof, Hanafiah and Ismail, Luthffi and Hanapiah, Fazah Akhtar and Mohamed, Salina and Piah, Hanizah Ali and Zahari, Nur Ismarrubie},
  booktitle={2012 IEEE 8th International Colloquium on Signal Processing and its Applications},
  pages={188--193},
  year={2012},
  organization={IEEE}
}

@article{mohebbi2020human,
  title={Human-robot interaction in rehabilitation and assistance: a review},
  author={Mohebbi, Abolfazl},
  journal={Current Robotics Reports},
  volume={1},
  number={3},
  pages={131--144},
  year={2020},
  publisher={Springer}
}

@article{leite2013influence,
  title={The influence of empathy in human--robot relations},
  author={Leite, Iolanda and Pereira, Andr{\'e} and Mascarenhas, Samuel and Martinho, Carlos and Prada, Rui and Paiva, Ana},
  journal={International journal of human-computer studies},
  volume={71},
  number={3},
  pages={250--260},
  year={2013},
  publisher={Elsevier}
}

@article{shin2018empathy,
  title={Empathy and embodied experience in virtual environment: To what extent can virtual reality stimulate empathy and embodied experience?},
  author={Shin, Donghee},
  journal={Computers in human behavior},
  volume={78},
  pages={64--73},
  year={2018},
  publisher={Elsevier}
}

@article{rueda2020virtual,
  title={Virtual reality and empathy enhancement: Ethical aspects},
  author={Rueda, Jon and Lara, Francisco},
  journal={Frontiers in Robotics and AI},
  volume={7},
  pages={506984},
  year={2020},
  publisher={Frontiers Media SA}
}

@article{lv2022application,
  title={The application of virtual reality technology in the efficiency optimisation of students' online interactive learning},
  author={Lv, Ninghua and Gong, Jingjing},
  journal={International Journal of Continuing Engineering Education and Life Long Learning},
  volume={32},
  number={1},
  pages={35--47},
  year={2022},
  publisher={Inderscience Publishers (IEL)}
}

@article{nakazawa2023augmented,
  title={Augmented reality-based affective training for improving care communication skill and empathy},
  author={Nakazawa, Atsushi and Iwamoto, Miyuki and Kurazume, Ryo and Nunoi, Masato and Kobayashi, Masaki and Honda, Miwako},
  journal={PloS one},
  volume={18},
  number={7},
  pages={e0288175},
  year={2023},
  publisher={Public Library of Science San Francisco, CA USA}
}

@inproceedings{billinghurst2014using,
  title={Using augmented reality to create empathic experiences},
  author={Billinghurst, Mark},
  booktitle={Proceedings of the 19th international conference on Intelligent User Interfaces},
  pages={5--6},
  year={2014}
}

@article{sorin2024large,
  title={Large language models and empathy: systematic review},
  author={Sorin, Vera and Brin, Dana and Barash, Yiftach and Konen, Eli and Charney, Alexander and Nadkarni, Girish and Klang, Eyal},
  journal={Journal of medical Internet research},
  volume={26},
  pages={e52597},
  year={2024},
  publisher={JMIR Publications Toronto, Canada}
}

@inproceedings{hasan2023sapien,
  title={SAPIEN: affective virtual agents powered by large language models},
  author={Hasan, Masum and Ozel, Cengiz and Potter, Sammy and Hoque, Ehsan},
  booktitle={2023 11th International Conference on Affective Computing and Intelligent Interaction Workshops and Demos (ACIIW)},
  pages={1--3},
  year={2023},
  organization={IEEE}
}

@inproceedings{shih2024empathy,
  title={Empathy-GPT: Leveraging Large Language Models to Enhance Emotional Empathy and User Engagement in Embodied Conversational Agents},
  author={Shih, Meng Ting and Hsu, Ming Yun and Lee, Sheng Cian},
  booktitle={Adjunct Proceedings of the 37th Annual ACM Symposium on User Interface Software and Technology},
  pages={1--3},
  year={2024}
}

@article{ghafurian2021improving,
  title={Improving humanness of virtual agents and users’ cooperation through emotions},
  author={Ghafurian, Moojan and Budnarain, Neil and Hoey, Jesse},
  journal={IEEE Transactions on Affective Computing},
  volume={14},
  number={2},
  pages={1461--1471},
  year={2021},
  publisher={IEEE}
}

@article{lisetti2013can,
  title={I can help you change! an empathic virtual agent delivers behavior change health interventions},
  author={Lisetti, Christine and Amini, Reza and Yasavur, Ugan and Rishe, Naphtali},
  journal={ACM Transactions on Management Information Systems (TMIS)},
  volume={4},
  number={4},
  pages={1--28},
  year={2013},
  publisher={ACM New York, NY, USA}
}

@article{parmar2022designing,
  title={Designing empathic virtual agents: manipulating animation, voice, rendering, and empathy to create persuasive agents},
  author={Parmar, Dhaval and Olafsson, Stefan and Utami, Dina and Murali, Prasanth and Bickmore, Timothy},
  journal={Autonomous agents and multi-agent systems},
  volume={36},
  number={1},
  pages={17},
  year={2022},
  publisher={Springer}
}

@inproceedings{plumer2024xr,
  title={XR Prototyping of Mixed Reality Visualizations: Compensating Interaction Latency for a Medical Imaging Robot},
  author={Pl{\"u}mer, Jan Hendrik and Yu, Kevin and Eck, Ulrich and Kalkofen, Denis and Steininger, Philipp and Navab, Nassir and Tatzgern, Markus},
  booktitle={2024 IEEE International Symposium on Mixed and Augmented Reality (ISMAR)},
  pages={1--10},
  year={2024},
  organization={IEEE}
}

@inproceedings{surve2024unrealthasc,
  title={UnRealTHASC--A Cyber-Physical XR Testbed for Underwater Real-Time Human Autonomous Systems Collaboration},
  author={Surve, Sushrut and Guo, Jia and Menezes, Jovan C and Tate, Connor and Jin, Yiting and Walker, Justin and Ferrari, Silvia},
  booktitle={2024 33rd IEEE International Conference on Robot and Human Interactive Communication (ROMAN)},
  pages={196--203},
  year={2024},
  organization={IEEE}
}

@inproceedings{mielke2025virtual,
  title={Virtual Studies, Real Results? Assessing the Impact of Virtualization on Human-Robot Interaction},
  author={Mielke, Tonia and Allgaier, Mareen and Schott, Danny and Hansen, Christian and Heinrich, Florian},
  booktitle={Proceedings of the Extended Abstracts of the CHI Conference on Human Factors in Computing Systems},
  pages={1--8},
  year={2025}
}

@inproceedings{stotko2019vr,
  title={A VR system for immersive teleoperation and live exploration with a mobile robot},
  author={Stotko, Patrick and Krumpen, Stefan and Schwarz, Max and Lenz, Christian and Behnke, Sven and Klein, Reinhard and Weinmann, Michael},
  booktitle={2019 IEEE/RSJ International Conference on Intelligent Robots and Systems (IROS)},
  pages={3630--3637},
  year={2019},
  organization={IEEE}
}

@inproceedings{li2019comparing,
  title={Comparing human-robot proxemics between virtual reality and the real world},
  author={Li, Rui and van Almkerk, Marc and van Waveren, Sanne and Carter, Elizabeth and Leite, Iolanda},
  booktitle={2019 14th ACM/IEEE international conference on human-robot interaction (HRI)},
  pages={431--439},
  year={2019},
  organization={IEEE}
}

@article{schmitt2018soft,
  title={Soft robots manufacturing: A review},
  author={Schmitt, Fran{\c{c}}ois and Piccin, Olivier and Barb{\'e}, Laurent and Bayle, Bernard},
  journal={Frontiers in Robotics and AI},
  volume={5},
  pages={84},
  year={2018},
  publisher={Frontiers Media SA}
}

@article{karabegovic2015application,
  title={The application of service robots for logistics in manufacturing processes.},
  author={Karabegovi{\'c}, I and Karabegovi{\'c}, E and Mahmi{\'c}, M and Husak, EJAIPE},
  journal={Advances in Production Engineering \& Management},
  volume={10},
  number={4},
  year={2015}
}

@article{ringwald2023should,
  title={How should your assistive robot look like? A scoping review on embodiment for assistive robots},
  author={Ringwald, Marina and Theben, Paulina and Gerlinger, Ken and Hedrich, Annika and Klein, Barbara},
  journal={Journal of Intelligent \& Robotic Systems},
  volume={107},
  number={1},
  pages={12},
  year={2023},
  publisher={Springer}
}

@article{willemse2017affective,
  title={Affective and behavioral responses to robot-initiated social touch: toward understanding the opportunities and limitations of physical contact in human--robot interaction},
  author={Willemse, Christian JAM and Toet, Alexander and Van Erp, Jan BF},
  journal={Frontiers in ICT},
  volume={4},
  pages={12},
  year={2017},
  publisher={Frontiers Media SA}
}

@incollection{trevelyan2016robotics,
  title={Robotics in hazardous applications},
  author={Trevelyan, James and Hamel, William R and Kang, Sung-Chul},
  booktitle={Springer handbook of robotics},
  pages={1521--1548},
  year={2016},
  publisher={Springer}
}

@article{kim2022effects,
  title={Effects on co-presence of a virtual human: A comparison of display and interaction types},
  author={Kim, Daehwan and Jo, Dongsik},
  journal={Electronics},
  volume={11},
  number={3},
  pages={367},
  year={2022},
  publisher={MDPI}
}

@article{lee2025conceptual,
  title={A conceptual model of immersive experience in extended reality},
  author={Lee, Hyunkook},
  journal={Computers in Human Behavior Reports},
  pages={100663},
  year={2025},
  publisher={Elsevier}
}

@article{awais2025foundation,
  title={Foundation models defining a new era in vision: a survey and outlook},
  author={Awais, Muhammad and Naseer, Muzammal and Khan, Salman and Anwer, Rao Muhammad and Cholakkal, Hisham and Shah, Mubarak and Yang, Ming-Hsuan and Khan, Fahad Shahbaz},
  journal={IEEE Transactions on Pattern Analysis and Machine Intelligence},
  year={2025},
  publisher={IEEE}
}

@article{lu2023mathvista,
  title={Mathvista: Evaluating mathematical reasoning of foundation models in visual contexts},
  author={Lu, Pan and Bansal, Hritik and Xia, Tony and Liu, Jiacheng and Li, Chunyuan and Hajishirzi, Hannaneh and Cheng, Hao and Chang, Kai-Wei and Galley, Michel and Gao, Jianfeng},
  journal={arXiv preprint arXiv:2310.02255},
  year={2023}
}

@article{li2024multimodal,
  title={Multimodal foundation models: From specialists to general-purpose assistants},
  author={Li, Chunyuan and Gan, Zhe and Yang, Zhengyuan and Yang, Jianwei and Li, Linjie and Wang, Lijuan and Gao, Jianfeng and others},
  journal={Foundations and Trends{\textregistered} in Computer Graphics and Vision},
  volume={16},
  number={1-2},
  pages={1--214},
  year={2024},
  publisher={Now Publishers, Inc.}
}

@incollection{kupcsik2017learning,
  title={Learning dynamic robot-to-human object handover from human feedback},
  author={Kupcsik, Andras and Hsu, David and Lee, Wee Sun},
  booktitle={Robotics Research: Volume 1},
  pages={161--176},
  year={2017},
  publisher={Springer}
}

@article{lin2020review,
  title={A review on interactive reinforcement learning from human social feedback},
  author={Lin, Jinying and Ma, Zhen and Gomez, Randy and Nakamura, Keisuke and He, Bo and Li, Guangliang},
  journal={IEEE Access},
  volume={8},
  pages={120757--120765},
  year={2020},
  publisher={IEEE}
}

@inproceedings{ruhland2015review,
  title={A review of eye gaze in virtual agents, social robotics and hci: Behaviour generation, user interaction and perception},
  author={Ruhland, Kerstin and Peters, Christopher E and Andrist, Sean and Badler, Jeremy B and Badler, Norman I and Gleicher, Michael and Mutlu, Bilge and McDonnell, Rachel},
  booktitle={Computer graphics forum},
  volume={34},
  number={6},
  pages={299--326},
  year={2015},
  organization={Wiley Online Library}
}

@article{qi2024computer,
  title={Computer vision-based hand gesture recognition for human-robot interaction: a review},
  author={Qi, Jing and Ma, Li and Cui, Zhenchao and Yu, Yushu},
  journal={Complex \& Intelligent Systems},
  volume={10},
  number={1},
  pages={1581--1606},
  year={2024},
  publisher={Springer}
}

@article{lynch2023interactive,
  title={Interactive language: Talking to robots in real time},
  author={Lynch, Corey and Wahid, Ayzaan and Tompson, Jonathan and Ding, Tianli and Betker, James and Baruch, Robert and Armstrong, Travis and Florence, Pete},
  journal={IEEE Robotics and Automation Letters},
  year={2023},
  publisher={IEEE}
}

@inproceedings{duguleana2011evaluating,
  title={Evaluating human-robot interaction during a manipulation experiment conducted in immersive virtual reality},
  author={Duguleana, Mihai and Barbuceanu, Florin Grigorie and Mogan, Gheorghe},
  booktitle={International Conference on Virtual and Mixed Reality},
  pages={164--173},
  year={2011},
  organization={Springer}
}

@inproceedings{bayro2022subjective,
  title={Subjective and objective analyses of collaboration and co-presence in a virtual reality remote environment},
  author={Bayro, Allison and Ghasemi, Yalda and Jeong, Heejin},
  booktitle={2022 ieee conference on virtual reality and 3d user interfaces abstracts and workshops (vrw)},
  pages={485--487},
  year={2022},
  organization={IEEE}
}

@inproceedings{zhang2023see,
  title={See or hear? exploring the effect of visual/audio hints and gaze-assisted instant post-task feedback for visual search tasks in ar},
  author={Zhang, Yuchong and Nowak, Adam and Xuan, Yueming and Romanowski, Andrzej and Fjeld, Morten},
  booktitle={2023 IEEE International Symposium on Mixed and Augmented Reality (ISMAR)},
  pages={1113--1122},
  year={2023},
  organization={IEEE}
}

@article{zhang2021supporting,
  title={Supporting visualization analysis in industrial process tomography by using augmented reality—a case study of an industrial microwave drying system},
  author={Zhang, Yuchong and Omrani, Adel and Yadav, Rahul and Fjeld, Morten},
  journal={Sensors},
  volume={21},
  number={19},
  pages={6515},
  year={2021},
  publisher={MDPI}
}

@inproceedings{nowak2021augmented,
  title={Augmented reality with industrial process tomography: to support complex data analysis in 3D space},
  author={Nowak, Adam and Zhang, Yuchong and Romanowski, Andrzej and Fjeld, Morten},
  booktitle={Adjunct Proceedings of the 2021 ACM International Joint Conference on Pervasive and Ubiquitous Computing and Proceedings of the 2021 ACM International Symposium on Wearable Computers},
  pages={56--58},
  year={2021}
}

@article{xia2024shaping,
  title={Shaping high-performance wearable robots for human motor and sensory reconstruction and enhancement},
  author={Xia, Haisheng and Zhang, Yuchong and Rajabi, Nona and Taleb, Farzaneh and Yang, Qunting and Kragic, Danica and Li, Zhijun},
  journal={Nature Communications},
  volume={15},
  number={1},
  pages={1760},
  year={2024},
  publisher={Nature Publishing Group UK London}
}

@inproceedings{zhang2024vision,
  title={Vision beyond boundaries: An initial design space of domain-specific large vision models in human-robot interaction},
  author={Zhang, Yuchong and Ma, Yong and Kragic, Danica},
  booktitle={Adjunct Proceedings of the 26th International Conference on Mobile Human-Computer Interaction},
  pages={1--8},
  year={2024}
}

@article{zhang2025mind,
  title={Mind meets robots: a review of EEG-based brain-robot interaction systems},
  author={Zhang, Yuchong and Rajabi, Nona and Taleb, Farzaneh and Matviienko, Andrii and Ma, Yong and Bj{\"o}rkman, M{\aa}rten and Kragic, Danica},
  journal={International Journal of Human--Computer Interaction},
  pages={1--32},
  year={2025},
  publisher={Taylor \& Francis}
}

@inproceedings{zhang2023playing,
  title={Playing with data: An augmented reality approach to interact with visualizations of industrial process tomography},
  author={Zhang, Yuchong and Xuan, Yueming and Yadav, Rahul and Omrani, Adel and Fjeld, Morten},
  booktitle={IFIP Conference on Human-Computer Interaction},
  pages={123--144},
  year={2023},
  organization={Springer}
}

@inproceedings{zhang2023industrial,
  title={Is industrial tomography ready for augmented reality? a need-finding study of how augmented reality can be adopted by industrial tomography experts},
  author={Zhang, Yuchong and Nowak, Adam and Rao, Guruprasad and Romanowski, Andrzej and Fjeld, Morten},
  booktitle={International Conference on Human-Computer Interaction},
  pages={523--535},
  year={2023},
  organization={Springer}
}

@inproceedings{zhang2022initial,
  title={An initial exploration of visual cues in head-mounted display augmented reality for book searching},
  author={Zhang, Yuchong and Nowak, Adam and Romanowski, Andrzej and Fjeld, Morten},
  booktitle={Proceedings of the 21st International Conference on Mobile and Ubiquitous Multimedia},
  pages={273--275},
  year={2022}
}

@inproceedings{zhang2022site,
  title={On-site or remote working?: An initial solution on how covid-19 pandemic may impact augmented reality users},
  author={Zhang, Yuchong and Nowak, Adam and Romanowski, Andrzej and Fjeld, Morten},
  booktitle={Proceedings of the 2022 International Conference on Advanced Visual Interfaces},
  pages={1--3},
  year={2022}
}

@inproceedings{wallace2024imitation,
  title={Imitation or innovation? translating features of expressive motion from humans to robots},
  author={Wallace, Benedikte and Van Otterdijk, Marieke and Zhang, Yuchong and Rajabi, Nona and Marin-Bucio, Diego and Kragic, Danica and Torresen, Jim},
  booktitle={Proceedings of the 12th International Conference on Human-Agent Interaction},
  pages={296--304},
  year={2024}
}

@inproceedings{baytas2019design,
  title={The design of social drones: A review of studies on autonomous flyers in inhabited environments},
  author={Baytas, Mehmet Aydin and {\c{C}}ay, Damla and Zhang, Yuchong and Obaid, Mohammad and Yanta{\c{c}}, Asim Evren and Fjeld, Morten},
  booktitle={Proceedings of the 2019 CHI Conference on Human Factors in Computing Systems},
  pages={1--13},
  year={2019}
}

@inproceedings{zhang2024human,
  title={Human-centered AI technologies in human-robot interaction for social settings},
  author={Zhang, Yuchong and Kassem, Khaled and Gong, Zhengya and Mo, Fan and Ma, Yong and Kirjavainen, Emma and H{\"a}kkil{\"a}, Jonna},
  booktitle={Proceedings of the International Conference on Mobile and Ubiquitous Multimedia},
  pages={501--505},
  year={2024}
}

\end{document}